\documentclass[multphys,vecphys]{svmult}

\usepackage{makeidx}         
\usepackage{graphicx}        
\usepackage{multicol}        
\usepackage{cite}            
\usepackage{amssymb}

\usepackage[bottom]{footmisc}

\makeindex             

\def\betasf{$\beta ''$-(BEDT-TTF)$_2$SF$_5$CH$_2$CF$_2$SO$_3$}
\def\cuscn{$\kappa$-(BEDT-TTF)$_2$Cu(NCS)$_2$}
\def\khg{$\alpha$-(BEDT-TTF)$_2$KHg(SCN)$_4$}

\def\mhg{$\alpha$-(BEDT-TTF)$_2M$Hg(SCN)$_4$}
\def\nh4{$\alpha$-(BEDT-TTF)$_2$NH$_4$Hg(SCN)$_4$}

\def\per{(Per)$_2M$(mnt)$_2$}
\def\pera{(Per)$_2$Au(mnt)$_2$}

\begin{document}

\title{High-field magnetoresistive effects in 
reduced-dimensionality organic metals and superconductors}
\titlerunning{Magnetoresistive effects in low-D
organic metals and superconductors}

\author{J. Singleton, R.D. McDonald \and N. Harrison.}
\institute{National High Magnetic Field 
Laboratory, LANL, MS-E536, Los Alamos, New Mexico 87545, USA 
\texttt{jsingle@lanl.gov; rmcd@lanl.gov; nharrison@lanl.gov}}

\maketitle
\begin{abstract}
The large charge-transfer anisotropy
of quasi-one- and quasi-two-dimensional
crystalline organic metals means that
magnetoresistance is one of the most powerful tools
for probing their bandstructure and interesting phase diagrams.
Here we review various magnetoresistance 
phenomena that are of interest in 
the investigation
of metallic, superconducting and charge-density-wave 
organic systems.
\end{abstract}


%

\section{Introduction}
\label{intro}
Quasi-two-dimensional crystalline organic metals and 
superconductors are very flexible systems in the study 
of many-body effects and unusual mechanisms for 
superconductivity~\cite{r0a,r0b,r1,r1a,r1b,r2,r3}.
Their ``soft'' lattices enable one to use 
relatively low pressures 
to tune the same material through a variety of 
low-temperature groundstates, for example from Mott 
insulator via intermingled 
antiferromagnetic and superconducting 
states to unusual superconductor~\cite{r1a,r2,r3}. 
Pressure also provides a sensitive means of varying the 
electron-phonon and electron-electron interactions, 
allowing their influence on the superconducting 
groundstate to be mapped~\cite{r1,r1a,r4}. 
The self-organising 
tendencies of organic molecules means that organic metals 
and superconductors are often rather clean and 
well-ordered systems; as we shall see below,
this enables the Fermi-surface 
topology to be measured in very great detail
using modest magnetic fields~\cite{r1,r5}.
Such information can then be used as input
parameters for theoretical models~\cite{r1}.
And yet the same organic molecules can adopt a variety 
of configurations, leading to ``glassy''
structural transitions and mixed phases in otherwise very pure 
systems~\cite{r1a,r6,r7}; these states may be important 
precursors to the superconductivity in such cases~\cite{r7}. 

Intriguingly, there seem to be at least two (or possibly three)
distinct mechanisms for superconductivity~\cite{r1,r8,r9,review}
in the quasi-two-dimensional organic conductors.
The first applies to half-filled-band
layered charge-transfer salts, such as the
$\kappa-$, $\beta-$ and $\beta'-$ 
packing arrangements of salts
of the form (BEDT-TTF)$_2$X,
where X is an anion molecule;
the superconductivity appears to be 
mediated by electron correlations/antiferromagnetic 
fluctuations~\cite{r1,r1a,r1b}.
The second mechanism applies to 
{\it e.g.} the $\beta''$ phase BEDT-TTF salts~\cite{r1a};
it appears to depend on the proximity of a metallic phase to 
charge order~\cite{r7,r8,r9}.
Finally, there may be some instances 
of BCS-like phonon-mediated
superconductivity~\cite{review}.

\begin{figure}[htbp]
\centering
\includegraphics[width=6cm]{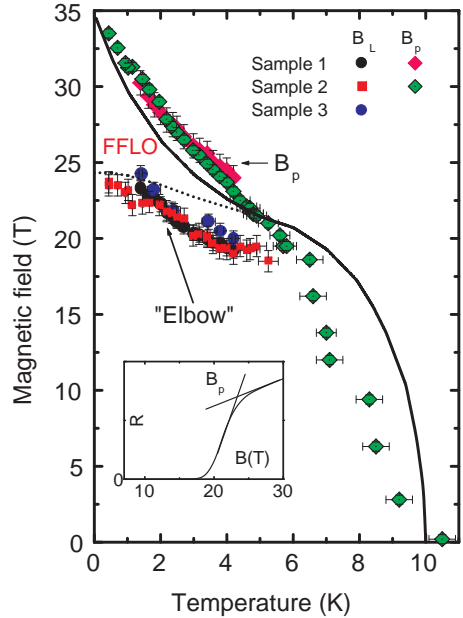}
\caption{Observation of the Fulde-Ferrell-Larkin-Ovchinnikov
(FFLO) phase in \cuscn ~\cite{r11}. The points labelled
$B_{\rm P}$ denote the resistive upper critical field
for two different samples (see inset for an 
illustrative example).
The $B_{\rm L}$ points, denoting
the phase boundary between the mixed
phase and FFLO state were deduced using simultaneous
MHz differential susceptibility measurements;
the change in vortex stiffness that accompanies
entry into the FFLO state causes an ``elbow''
in the field-dependent susceptibility.
``Sample 3'' is a measurement on a third sample
under different conditions of electric field;
consistency of the phase boundaries for the
three samples shows that the effect is not
due to artefacts of vortex pinning. The
curves are a theoretical model
due to Shimahara for the 
upper critical field and FFLO
(see Ref.~\cite{r11} for details).}
\label{fig1}
\end{figure}

The main purpose of this chapter is to discuss
the role that high magnetic fields and
magnetoresistance measurements, can play in unravelling
the above-mentioned properties of quasi-one-
and quasi-two-dimensional organic metals
and superconductors.
Hence, we shall spend some time discussing
the high-field magnetotransport experiments that have helped 
to measure the Fermi surfaces of charge-transfer salts 
of molecules such as BEDT-TTF and BETS.
In addition to their invaluable role in mapping the
bandstructure, high magnetic fields
allow one to tune some of the organic
conductors into some new and intriguing phases;
magnetoresistance phenomena can then be used to
delineate the phase boundary of the new state.
Examples include field-induced superconductivity~\cite{r10}
and exotic states such as the Fulde-Ferrell-Larkin-Ovchinnikov 
(FFLO) phase~\cite{r11}. The phase diagram of the
latter state in \cuscn ~is shown in Fig.~\ref{fig1};
its derivation is a good illustration of the 
general utility of high fields
and magnetoresistive phenomena.
First, a conventional ($\sim 30$~Hz)
measurement of the magnetoresistance
was used to very precisely orient the sample in the magnetic
field and to measure the superconducting-to-resistive
transition~\cite{r11}. Subsequently, high-frequency (MHz)
magnetoresistance measurements that are sensitive to changes
in dissipation {\it within} the zero-resistance state,
allowed the FFLO to type~II superconductivity
boundary to be measured~\cite{r11}.
In view of recent doubts about the 
proposed FFLO state in CeCoIn$_5$~\cite{doubts},
organic conductors such as \cuscn ~\cite{r11}
and $\lambda$-(BETS)$_2$GaCl$_4$~\cite{tanatar}
are perhaps as yet the only systems
in which the FFLO has been truly observed. 
Later in this paper, 
we shall describe other recent observations of 
field-induced phases in crystalline organic metals, 
which are related to the FFLO but which result in 
insulating states.

The remainder of this chapter is organized as follows.
Section~\ref{s1p2} describes the
Fermi-surface topologies of some typical
organic charge-transfer salts, concentrating on
the features that influence the magnetoresistance
in high fields; a brief mention is also made of the
deficiencies of bandstructure calculations.
Magnetoresistive phenomena are discussed
in Section~\ref{s1p3}, including measurements
of the effective dimensionality, angle-dependent
magnetoresistance oscillations (AMROs), 
the magnetoresistivity tensor and Fermi-surface-traversal
resonances. The Shubnikov-de Haas effect is treated in
Section~\ref{s1p4}, with a focus on the effects
of reduced dimensionality and the extraction of
quantities such as the 
quasiparticle scattering rate and effective mass.
Sections~\ref{s1p5} and \ref{s1p6} discuss
some of the phenomena associated with charge-density
waves above the Pauli paramagnetic limit.
Finally, some thoughts about future prospects are given in
Section~\ref{s1p7}

\section{Intralayer Fermi-surface topologies}
\label{s1p2}
The defining property of a metal is that it possesses
a {\it Fermi surface}, that is, a constant-energy
surface in
$k$-space which separates the filled electron
states from empty electron states at absolute zero ($T=0$).
The shape of the Fermi surface is determined
by the dispersion relationships (energy versus
{\bf k} relationships) $E=E({\bf k})$ of each
partially-filled band and the number of
quasiparticles to be accommodated.
As is mentioned elsewhere in this book, the crystal structures
of the organic metals and superconductors that feature in this
article are mostly layered (or chain-like), 
with planes of anions (and perhaps other
space-filling molecules~\cite{r7,r9a})
alternating with layers of the cation
molecules whose overlapping molecular orbitals provide the
electronic bands. The main consequence of this structural anisotropy is
that the intralayer (or interchain) transfer 
integrals tend to be much greater by a
factor $\sim 10^2-10^3$ than the interlayer ones,
so that most of the quasiparticle dispersion occurs 
parallel to the cation planes (or chains).
Consequently, for many experiments, including the Shubnikov-de Haas
and de Haas-van Alphen effects, the properties of the Fermi surface appear
almost exactly two dimensional. We shall return
to the consequences of this fact in later sections.

\begin{figure}[htbp]
\centering
\includegraphics[width=9cm]{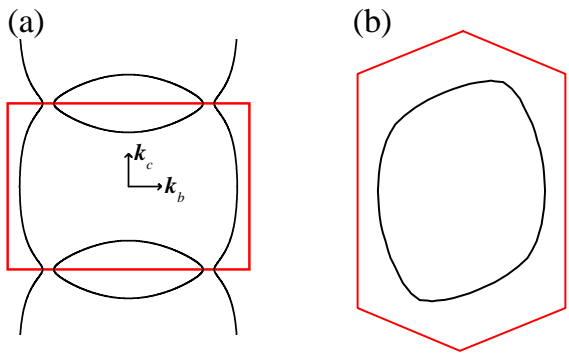}
\caption{(a)~Brillouin zone and Fermi-surface of
$\kappa$-(BEDT-TTF)$_2$Cu(NCS)$_2$,
showing the open, quasi-one-dimensional sections,
and the closed, quasi-two-dimensional
pocket~\cite{schmalian,caulfield,goddard}. 
(b)~Brillouin zone and Fermi-surface of
$\beta$-(BEDT-TTF)$_2$IBr$_2$~\cite{kartsovnik}.
}
\label{fspic}
\end{figure}

Figure~\ref{fspic} shows sections
(parallel to the highly-conducting planes) through the
first Brillouin zone and Fermi surfaces of two typical BEDT-TTF salts.
The example in Figure~\ref{fspic}(a) is
$\kappa$-(BEDT-TTF)$_2$Cu(NCS)$_2$~\cite{schmalian,caulfield,goddard};
$\kappa$-phase BEDT-TTF salts have a dimerized arrangement of BEDT-TTF
molecules such that there are four per unit cell, each pair (or dimer)
jointly donating one hole. 
The overall Fermi-surface cross-section is therefore
the same as that of the Brillouin zone.
However, the Fermi surface intersects
the Brillouin zone boundaries in the {\bf c} direction,
so that band gaps open up (see e.g. Chapter 2 of
Reference~\cite{singlebook}).
The Fermi surface thus splits into open (electron-like)
sections (often known as {\it Fermi sheets}) running down two of the
Brillouin-zone edges and a closed hole pocket (referred to as
the ``$\alpha$ pocket'')
straddling the other;
it is customary to label such sections
``quasi-one-dimensional'' and
``quasi-two-dimensional'' respectively.
The names arise because the group velocity
${\bf v}$ of the electrons is given
by~\cite{singlebook,ashcroft}
\begin{equation}
\hbar {\bf v} = \nabla_{\bf k} E({\bf k}).
\label{velocities}
\end{equation}
The Fermi surface is a surface of constant
energy; Equation~\ref{velocities}
shows that the velocities of electrons at the
Fermi surface will be directed perpendicular to it.
Therefore, referring to Figure~\ref{fspic},
electrons on the closed Fermi-surface pocket
can possess velocities which point in
any direction in the ($k_b,k_c$) plane;
they have freedom of movement
in two dimensions and are said to be
{\it quasi-two-dimensional}.
By contrast, electrons on the open sections
have velocities predominently directed parallel
to $k_b$ and are {\it quasi-one-dimensional}.
$\kappa$-phase BEDT-TTF superconductors $\kappa$-(BEDT-TTF)$_2X$
can be made with a variety
of other anion molecules, including
$X=$ Cu[N(CN)$_2$]Br (11.8~K), Cu[N(CN)$_2$]Cl (12.8~K (under pressure)),
and I$_3$ (4~K); here the number in parentheses represents $T_{\rm c}$.
In all of these salts, the Fermi-surface topology is very similar
to that in Figure~\ref{fspic}(a); small differences in the symmetry
of the anion layer lead to variations
in the gap between the quasi-one-dimensional and quasi-two-dimensional
Fermi-surface sections~\cite{review,mori,ishiguro}.
A summary of the detailed differences and effective
is given in Section 3.2 of
Reference~\cite{review} (see also \cite{mori}).
We shall see below (Section~\ref{s1p3p3}) that
magnetic breakdown~\cite{review}, in which 
the field-induced motion of quasiparticles
causes them to tunnel across the gaps 
between the Fermi-surface sections, leading to
new magnetic quantum oscillation frequencies,
is a common phenomenon in these salts~\cite{harrisonbd}.

Figure~\ref{fspic}(b) shows the Fermi-surface topology and
Brillouin zone of $\beta$-(BEDT-TTF)$_2$IBr$_2$~\cite{kartsovnik}.
In this case (see Figure~\ref{fspic}b) there is one hole
per unit cell, so that the Fermi surface cross-sectional area
is half that of the Brillouin zone; only a
quasi-two-dimensional pocket is present.

As mentioned above, the bandstructure
of a charge-transfer salt is chiefly determined by the
packing arrangement of the cation molecules.
The $\beta$, $\kappa$, $\beta ''$, $\lambda$ and $\alpha$
phases have tended to be the most commonly studied.
The latter four phases all have predictable Fermi surfaces
consisting of a quasi-two-dimensional pocket plus a 
pair of quasi-one-dimensional Fermi sheets
(the pocket arrangement differs from phase 
to phase)~\cite{review,wosnitza,mori};
the $\beta$-phase is alone in possessing a Fermi surface
consisting of a single quasi-two-dimensional 
pocket~\cite{ishiguro}.

As much of the rest of this article will be 
about using magnetoresistance to measure Fermi surfaces,
it is worth including a brief note on the deficiencies of 
bandstructure calculations habitually applied to the organics.
The bandstructures of crystalline organic metals have usually
been calculated using the extended 
H\"{u}ckel (tight-binding)\footnote{Simple
introductions to the tight-binding model of bandstructure
are given in References~\cite{singlebook,ashcroft}.} approach,
which employs the highest occupied molecular orbitals (HOMOs) of the
cation molecule~\cite{mori}. Section~5.1.3 of Reference~\cite{ishiguro}
discusses this approach and cites some of the most relevant papers.
Whilst this method is usually quite successful in predicting the main
features of the Fermi surface ({\it e.g.} the fact that
there are quasi-one-dimensional and 
quasi-two-dimensional Fermi-surface sections), the details
of the Fermi-surface topology are sometimes
inadequately described (see {\it e.g.}~\cite{eva}).
This can be important when, for example, the detailed corrugations
of a Fermi sheet govern the interactions
which determine its low-temperature groundstate~\cite{schmalian,eva}.
A possible way around this difficulty is to make slight adjustments of the
transfer integrals so that the predicted Fermi surface is in good
agreement with experimental 
measurements~\cite{schmalian,caulfield,goddard,eva}.
In the $\beta''$ and $\lambda$ phases the predicted bandstructure
seems very sensitive to the choice of basis set, and
the disagreement between calculation and measurement is often most severe
(see {\it e.g.}~\cite{betawos,doporto1,house,mielke}).

More sophisticated Hubbard-unrestricted Hartree-Fock band calculations
have been carried out for \cuscn ~\cite{demiralp}.
These calculations attempt to take into account many-body effects,
and are successful in reproducing a number of experimental
properties. They also indicate the importance of both
antiferromagnetic fluctuations and electron-phonon interactions
in \cuscn , a fact important in the proposed mechanisms
for supercoductivity~\cite{r1a,review}.
More recently, techniques such as 
Dynamical Mean Field Theory (DMFT) 
have been applied to quasi-two-dimensional
organic superconductors~\cite{r1a,newschmalian05,gabikotliar},
predicting some aspects of the complex phase diagram of the 
$\kappa$-phase BEDT-TTF salts.

\section{High-field magnetotransport effects}
\label{s1p3}
\subsection{Measurements of the effective 
Fermi-surface dimensionality via the SQUIT peak}
\label{s1p3p1}
We remarked above that the electronic properties
of quasi-two-dimensional organic metals are very anisotropic.
Many band-structure-measuring techniques chiefly
give information about the intralayer topology of the Fermi surface.
However, it is important to ask whether the Fermi surface is
exactly two-dimensional, or whether it extends in the interlayer direction,
i.e. is three-dimensional.

This question is of quite general interest, as
there are many correlated-electron systems which
have very anisotropic electronic bandstructure.
In addition to the organic superconductors~\cite{review,strong},
examples include the ``high-$T_{\rm c}$'' cuprates~\cite{cuprates},
and layered ruthenates~\cite{ruthenate} and manganites~\cite{ramirez}.
Such systems may be described by a tight-binding Hamiltonian
in which the ratio of the interlayer transfer integral $t_{\perp}$
to the average intralayer transfer integral $t_{||}$ is $\ll
1$~\cite{review,strong,mck}.
The inequality $\hbar/ \tau > t_{\perp}$~\cite{mott}
where $\tau^{-1}$ is the quasiparticle
scattering rate~\cite{cuprates,strong,mck},
frequently applies to such systems, suggesting that
the quasiparticles scatter more frequently than they tunnel between layers.
Similarly, under standard laboratory conditions, the inequality
$k_{\rm B}T > t_{\perp}$ often holds, hinting that
thermal smearing will ``wipe out'' 
details of the interlayer periodicity~\cite{anderson}.

The question has thus arisen as to whether the interlayer charge
transfer is coherent or incoherent in these materials,
i.e. whether or not the Fermi surface 
is a three-dimensional entity extending in the interlayer
direction~\cite{review,strong,mck}.
Incoherent interlayer transport is used as a justification
for a number of theories which are thought
to be pivotal in the understanding of
reduced-dimensionality materials (see e.g. \cite{strong,anderson}).
Moreover, models for unconventional superconductivity in $\kappa$-phase
BEDT-TTF salts invoke the nesting
properties of the
Fermi surface~\cite{schmalian,aoki,charffi};
the degree of nesting might depend on whether
the Fermi surface is two dimensional or three dimensional
(see \cite{review},
Section 3.5).

\begin{figure}[htbp]
\centering
\includegraphics[width=12cm]{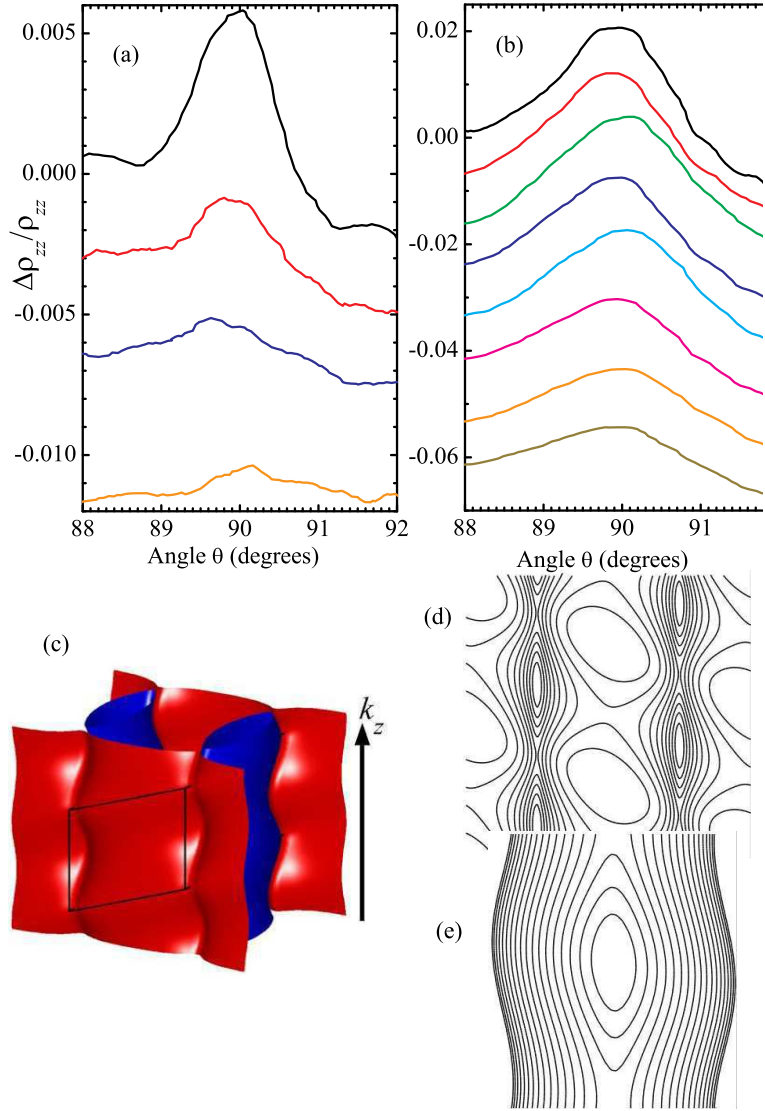}
\caption{Illustration of the ``SQUIT'' or
``coherence peak'' for in-plane fields
in \cuscn ~(after Ref.~\cite{goddardnew}). 
(a)~The 45~T magnetoresistance of \cuscn ~close to 
a tilt angle $\theta=90^{\circ}$
plotted as $\Delta \rho_{zz}/\rho_{zz{\rm BG}}$, the
fractional change in $\rho_{zz}$ from the more slowly-varying
background. Data for temperatures
$T=5.3$ (highest), 7.6, 9.6 and 13.1~K (lowest)
are shown, offset for visibility. 
In this plane of rotation, closed orbits
on the quasi-one-dimensional Fermi-surface sections (see (d) below)
are responsible for the SQUIT, observed as
a peak at $\theta=90^{\circ}$. (b)~Similar data for
a plane of rotation in which the SQUIT is
caused by closed orbits on the quasi-two-dimensional
Fermi-surface sections (see (e) below); 
the traces are for $T=5.3$ (highest), 
7.6, 8.6, 9.6, 10.6, 12.1, 13.1 
and 14.6~K (lowest). Each trace has been
offset for clarity. (c)~Three-dimensional 
representation of the Fermi surface of
\cuscn (after Ref.~\cite{goddardprb2004}); the finite
interlayer transfer integral 
gives the corrugations (shown greatly
exaggerated) on the sides of the FS. 
Quasi-one-dimensional and and quasi-two-dimensional 
Fermi-surface sections are shown in red and blue
respectively. (d)~Consequent
field-induced closed orbits on the side of the
quasi-one-dimensional Fermi-surface sections
section when $\theta=90^{\circ}$ and the field {\bf B}
is parallel to ${\bf k_b}$. (e)~Similar closed
orbits on the quasi-two-dimensional section when
$\theta=90^{\circ}$ and {\bf B} is
parallel to ${\bf k_c}$ (see Fig.~\ref{fspic}(a)). 
Orbits such as those in
(d) and (e) give rise to the SQUIT peak in $\rho_{zz}$ 
(see (a) and (b)).}
\label{belly}
\end{figure}

In this context, the experimental 
situation is at first sight complicated because
many apparently solid experimental tests for coherence
in organic metals and superconductors have been 
deemed to be inconclusive~\cite{mck};
e.g. semiclassical
models can reproduce AMRO~\cite{amro} 
and FTR data~\cite{schrama,edwardssingleton}
equally well when the interlayer transport
is coherent or ``weakly coherent''~\cite{mck}.
However, it turns out that magnetoresistance
{\it can} yield an unambiguous measurement 
of interlayer coherence,
by way of a feature in the interlayer
magnetoresistivity $\rho_{zz}$
known as the {\it SQUIT}
(Suppression of QUasiparticle Interlayer Transport)
or {\it coherence peak},
observed for exactly in-plane
fields (Figs.~\ref{belly}(a) and (b)).

To see how this comes about, consider
a simple tight-binding
expression for the interlayer ($z$-direction)
dispersion~\cite{review,goddardprb2004};
$E(k_z)=-2t_{\perp}\cos (k_za)$.
Here $t_{\perp}$ is the interlayer
transfer integral and $a$ is the unit-cell height in the $z$ direction.
The introduction of such an interlayer dispersion,
paired with an in-plane two-dimensional dispersion relationship, will
result in a three-dimensional Fermi surface with a 
sinusoidally-modulated Fermi-surface cross-section 
(see Ref.~\cite{singlebook}, Chapter 8).
A more realistic version of the same
idea is shown schematically for \cuscn ~in 
Figure~\ref{belly}(c)~\cite{footnote,cokebottle,disclaim} (compare
Fig.~\ref{fspic}(a))
If the Fermi surface is extended in the interlayer 
direction, a magnetic field
applied exactly in the intralayer plane
can cause a variety of orbits on the sides of the Fermi surface
(shown schematically in Fig.\ref{belly}(d) and (e)) via
the Lorentz force
\begin{equation}
\hbar({\rm d}{\bf k}/{\rm d}t)=-e{\bf v} \times {\bf B}
\end{equation}
where {\bf v} is given by Equation~\ref{velocities};
this results in orbits on the Fermi surface,
in a plane perpendicular to {\bf B}~\cite{review,ashcroft}.
Numerical modelling using a Chambers equation
approach and a realistic parameterization of
the Fermi surface~\cite{goddardprb2004}
show that the closed orbits
about the belly of the Fermi surface
are very effective in averaging $v_{\perp}$,
the interlayer component of the velocity.
Therefore, the presence of such orbits
will lead to an increase in the resistivity
component $\rho_{zz}$~\cite{goddardprb2004,hanasaki,russian}.
On tilting {\bf B} away from the intralayer direction,
the closed orbits cease to be possible when
{\bf B} has turned through an angle $\Delta$,
where~\cite{goddardprb2004}
\begin{equation}
\Delta ({\rm in~radians}) \approx v_{\perp}/v_{||}.
\label{delta}
\end{equation}
Here $v_{\perp}$ is the maximum of the interlayer
component of the quasiparticle velocity, and $v_{||}$ is
the intralayer component of the quasiparticle velocity
in the plane of rotation of {\bf B}.
Therefore, on tilting {\bf B} through the in-plane orientation,
one expects to see a peak in $\rho_{zz}$, of angular width $2 \Delta$,
if (and only if~\cite{mck}) the Fermi surface 
is extended in the $z$ direction.
It is this peak that is referred to as the ``coherence peak"
or the ``SQUIT'' peak (Figs.~\ref{belly}(a) and (b)).
By using Equations~\ref{velocities} and \ref{delta}
and measured details of the
intralayer Fermi-surface topology, it is possible to
use $\Delta$ to deduce $t_{\perp}$~\cite{goddardprb2004}
to considerable accuracy.

Figures~\ref{belly}(a) and (b) show typical data for \cuscn .
The observation of a peak in
$\rho_{zz}$ close to $\theta =90^{\circ}$ allows the
interlayer transfer integral to be estimated to be
$t_{\perp} \approx 0.065$~meV~\cite{goddardprb2004}.
This may be compared with intralayer transfer 
integrals $\sim 150$~meV~\cite{caulfield}.

Such data are of great interest because they illustrate that the
criteria frequenctly
used to delineate interlayer incoherence are rather a poor
guide to reality. For example, a temperature of
$T\approx 15$~K, ($k_{\rm B}T \approx 30 t_{\perp}$)
leads one to expect
incoherent interlayer transport via the criteron $k_{\rm B}T > t_{\perp}$
proposed by Anderson~\cite{anderson},
yet the peak in $\rho_{zz}$ shown in Figure~\ref{belly}(b) unambiguously
demonstrates a three-dimensional Fermi-surface topology~\cite{goddardnew}.

Demonstrations of interlayer coherence have been carried out on the 
quasi-two-dimensional organic
conductors $\beta$-(BEDT-TTF)$_2$IBr$_2$~\cite{kartsovnik},
$\kappa$-(BEDT-TTF)$_2$Cu$_2$(CN)$_3$~\cite{ohmichi} (under pressure),
\nh4 ~\cite{hanasaki} (under pressure), 
$\beta$-(BEDT-TTF)$_2$I$_3$~\cite{hanasaki},
\cuscn ~\cite{goddard},
$\lambda$-(BETS)$_2$GaCl$_4$ ~\cite{mielke} and
 \betasf ~\cite{wosnisquit,janesquit}.
In the latter example, no peak was observed, suggesting incoherent
interlayer transport, or no warping.
In all of the other instances, the $\rho_{zz}$ data demonstrate a
Fermi surface which is extended in the interlayer direction.

Inspired by this success, the technique has 
recently been extended to systems such as cuprate
superconductors~\cite{hussey}.

\subsection{Mechanisms for angle-dependent magnetoresistance oscillations
in quasi-two-dimensional organic metals}
\label{s1p3p2}
Whilst they give very accurate information about the cross-sectional
areas of the Fermi-surface sections, magnetic quantum oscillations
do not provide any details of their {\it shape} 
(see Section~\ref{s1p4}).
Such information is usually derived from 
angle-dependent magnetoresistance
oscillations (AMROs)~\cite{review,wosnitza,amro,goddardprb2004,amro2}.
AMROs are measured by rotating a 
sample in a fixed magnetic field
whilst monitoring its resistance; the coordinate used to denote
the position of AMROs is the polar angle $\theta$ between the
normal to the sample's quasi-two-dimensional 
planes and the magnetic field~\cite{amro,amro2}.
It is also very informative to vary the plane of rotation of
the sample in the field; this is 
described by the azimuthal angle 
$\phi$~\cite{amro,goddardprb2004,goddardnew,amro2}).

As has been mentioned in the previous section,
many quasi-two-dimensional charge-transfer salts of molecules such as
BEDT-TTF and BETS exhibit the SQUIT or coherence peak, showing that
they possess a well-defined three-dimensional 
Fermi surface, even at quite
elevated temperatures~\cite{goddardprb2004}. 
In such cases, the AMROs can be modeled 
using a Boltzmann-transport approach which 
treats the time-evolution of quasiparticle 
velocities across a three-dimensional Fermi 
surface\footnote{As mentioned elsewhere in this book,
the picture for TMTSF salts can be rather different.}.
 
\begin{figure}[htbp]
\centering
\includegraphics[height=7cm]{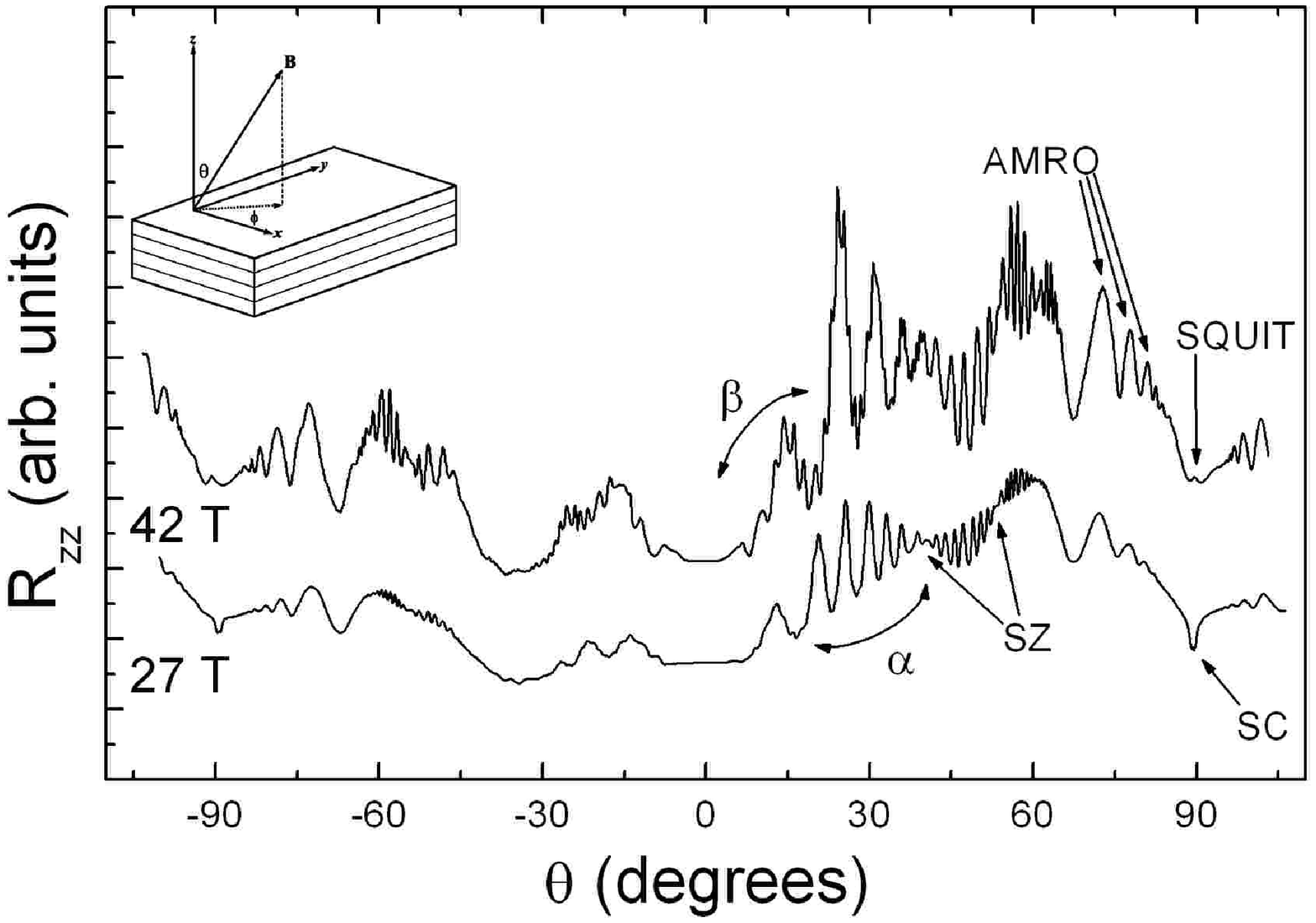}
\includegraphics[height=9cm]{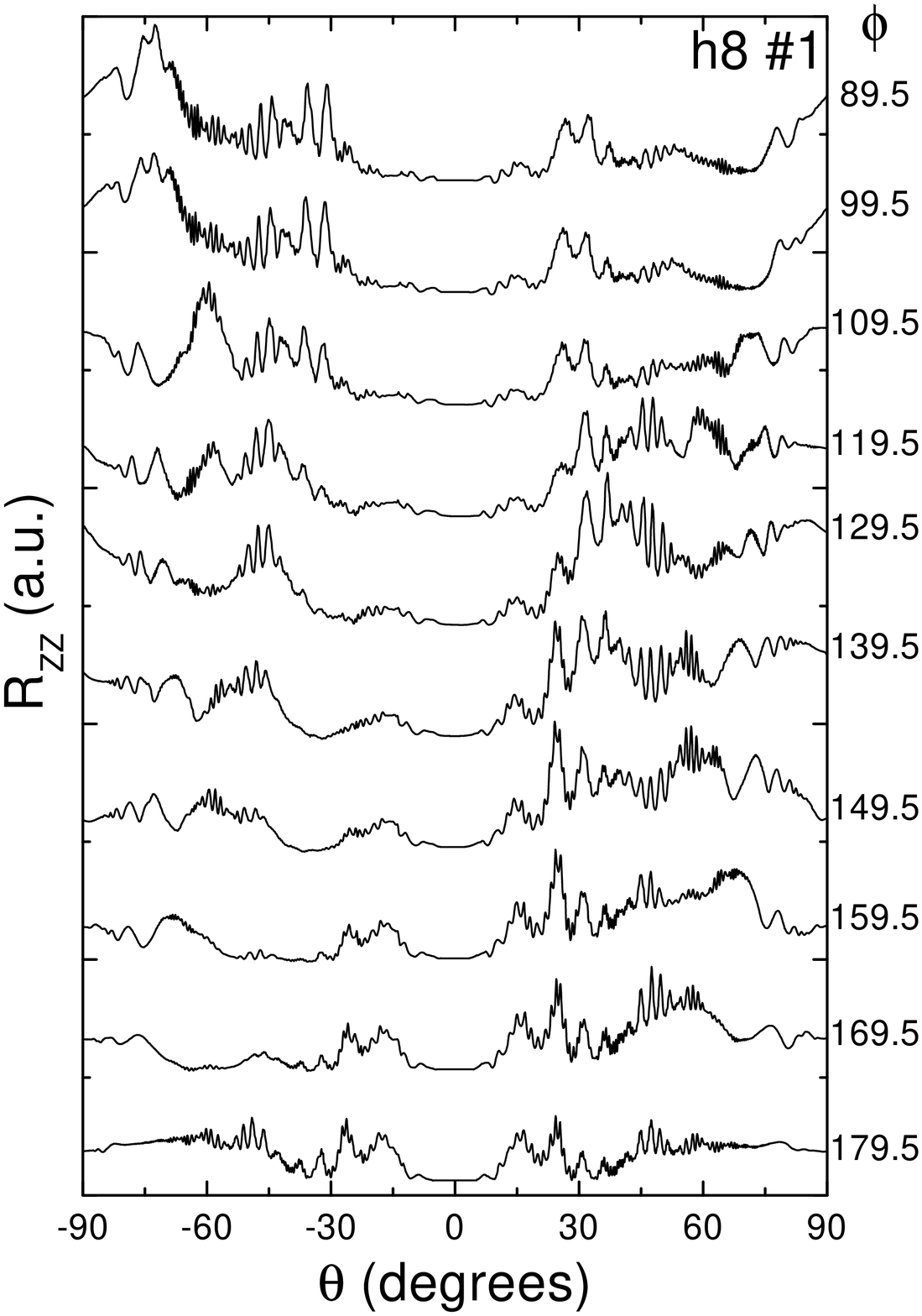}
\caption{(a) Typical $\theta$ dependence of the 
magnetoresistance of \cuscn ~\cite{goddardprb2004}. 
The data 
shown are for a hydrogenated sample at 490 mK,
$\phi = 149\deg$ (where $\phi$ is the azimuthal angle) 
and fields of 27 T (lower), and 42 T (upper). 
The data have been offset for clarity. Some representative features
are indicated; Shubnikov-de Haas (SdH) oscillations due to 
the Q2D pockets ($\alpha$) and the breakdown orbit 
($\beta$); spin zeros in the SdH amplitudes (SZ); the onset
of the superconducting transition (SC); 
angle-dependent magnetoresistance oscillations (AMRO), 
whose positions are field independent; and
the resistive SQUIT peak in the presence of an exactly 
in-plane magnetic field (in-plane Peak). 
The inset diagram is included to illustrate the
measurement geometry.  
(b) The angle-dependent interlayer magnetoresistance of
the same sample  at various values of the azimuthal angle $\phi$;
$T=500$~mK and $B=42$~T.}
\label{amros}
\end{figure}

\begin{figure}[htbp]
\centering
\includegraphics[width=9cm]{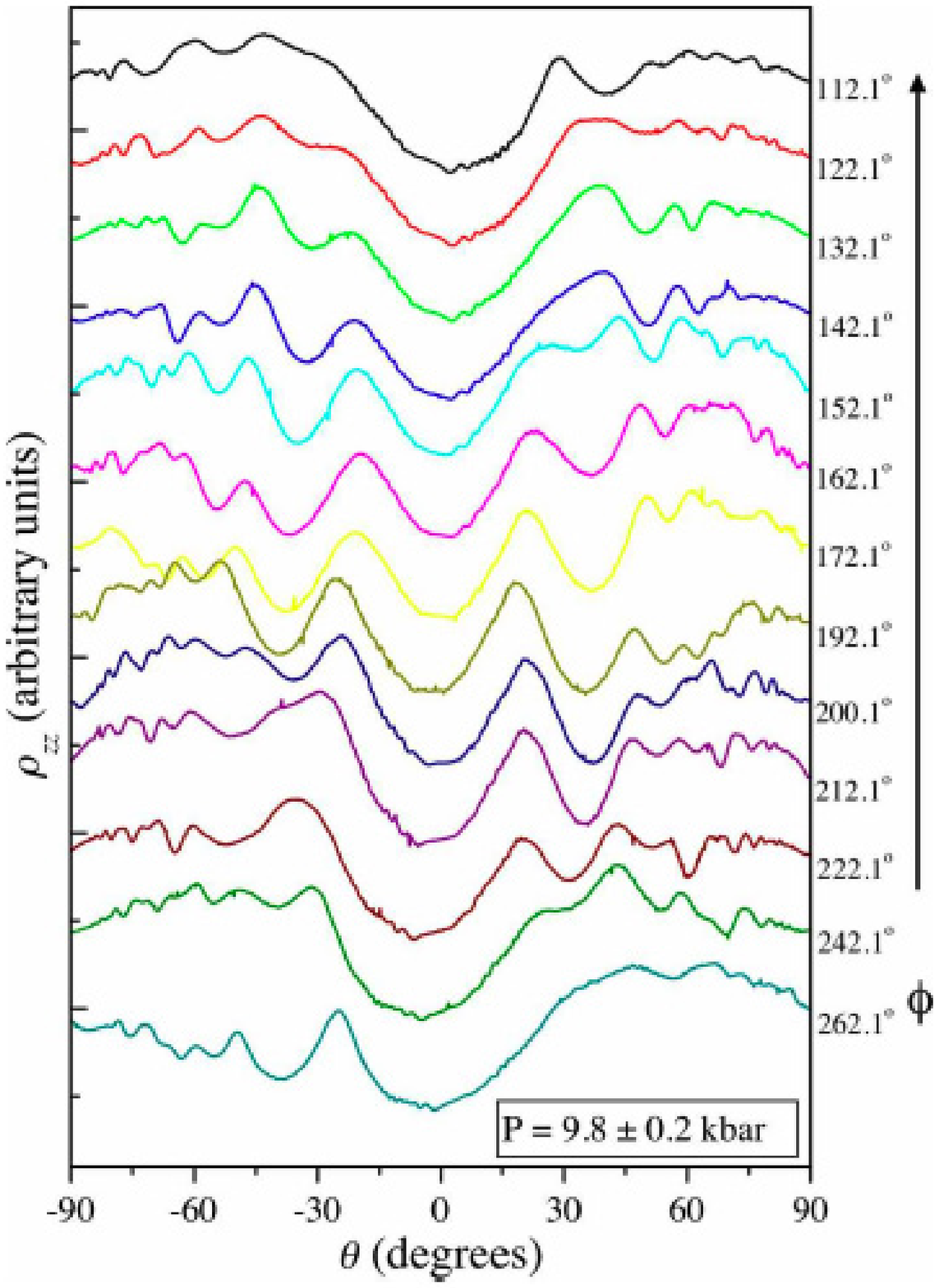}
\includegraphics[width=9cm]{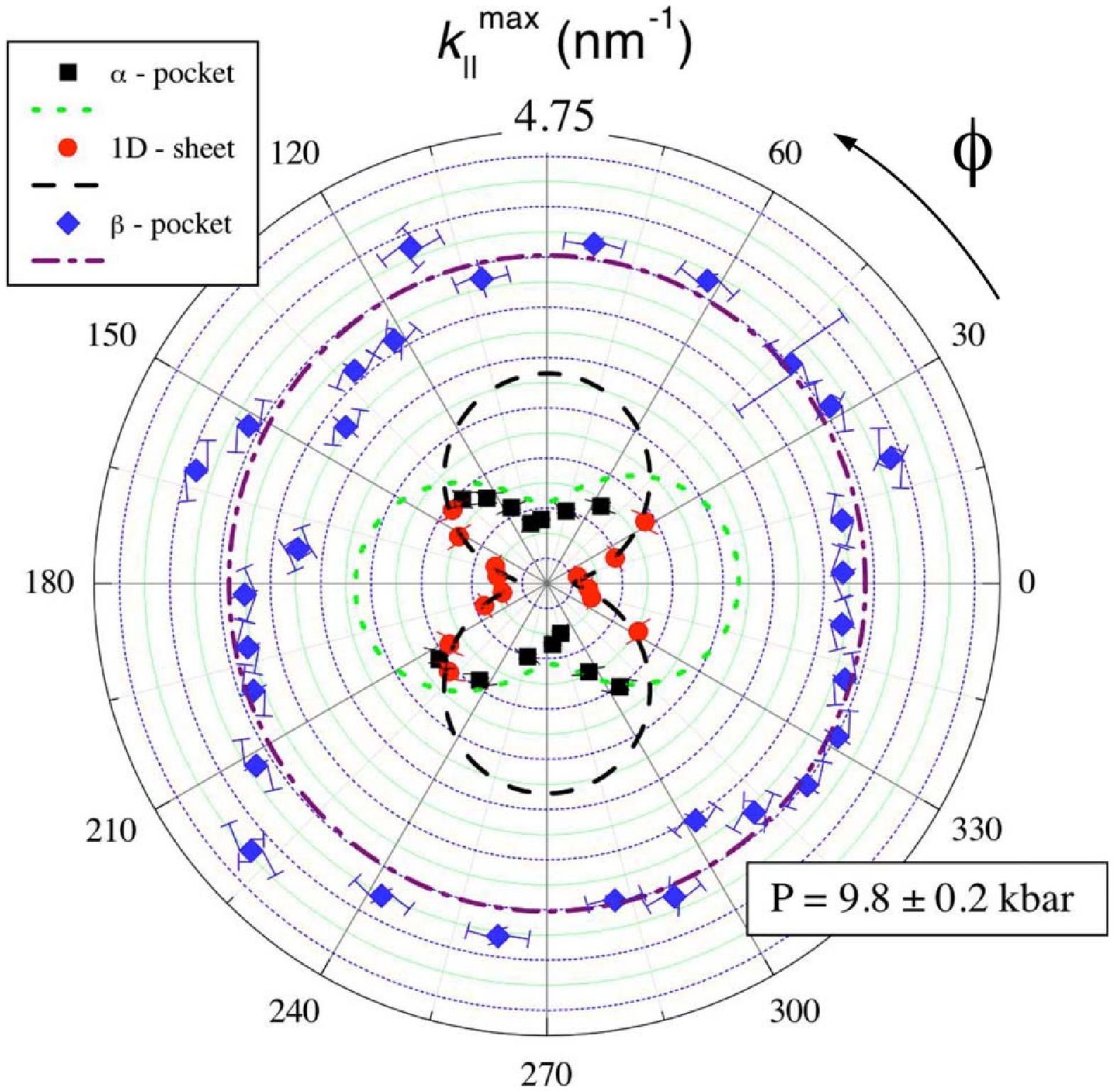}
\caption{Top: typical angle-dependent magnetoresistance 
oscillations (AMROs) at a magnetic 
field of 30~T and a temperature of 1.5~K; the pressure 
is 9.8~kbar. Here, $\theta$
denotes the angle between the normal to the 
sample's quasi-two-dimensional planes and the field; 
$\phi$ describes the plane of rotation. 
Bottom: polar plot of the periodicities (in units of tan $\theta$) 
of the various AMRO series. 
The inset key gives the mechanism for the features, 
with the blue diamonds representing the oscillations 
due to magnetic breakdown~\cite{bangura}.
 }
\label{bmros}
\end{figure}

In such a picture, AMROs result from the 
averaging effect that the semiclassical orbits
on the Fermi surface have on the quasiparticle velocity.
Both quasi-one-dimensional and quasi-two-dimensional 
Fermi-surface sections can give rise to AMROs;
in the former case, the AMROs are sharp dips in the resistivity,
periodic in $\tan \theta$;
in the latter case, one expects {\it peaks}, 
also periodic in $\tan \theta$~\cite{amro,amro2}.
In order to distinguish between these two cases, 
it is necessary to carry out
the experiment at several different $\phi$ 
(some fine cautionary hints are
given in Ref.~\cite{goddardprb2004}).
The $\phi$-dependence of the AMROs can be related directly to the
shape of a quasi-two-dimensional Fermi-surface section; 
in the case of a quasi-one-dimensional sheet,
the AMROs yield precise information about the sheet's
orientation~\cite{review,wosnitza,amro,goddardprb2004,amro2}.
Typical data are shown in Figure~\ref{amros}. Numerical
modelling of such data (using a Boltzmann transport approach)
allows a detailed three-dimensional picture of the Fermi surface to
be built up (see Figure~\ref{bmros}(a)).

A third type of AMRO has been observed
in the angle-dependent magnetoresistance of 
\cuscn ~subjected to high pressures.
The experiments employ a miniature diamond-anvil cell, 
attached to a cryogenic goniometer,
providing full two axis rotation at $^3$He temperatures~\cite{bangura}.
The apparatus is placed in a 33 T Bitter magnet. A plethora 
of AMROs is observed at each pressure 
(Figure~\ref{bmros}(b)), caused by field-induced 
quasiparticle orbits across the Fermi surface. 
Raising pressure suppresses the gap between the 
quasi-two-dimensional pocket and quasi-one-dimensional 
sheets of the Fermi surface (see Figure~\ref{fspic}(a)), 
increasing the probability 
of magnetic breakdown (see Section~\ref{s1p3p3}
and Ref.~\cite{harrisonbd}
for a more detailed explanation of magnetic breakdown). 
This permits AMROs 
due to breakdown orbits about the 
complete Fermi surface. 

\begin{figure}[htbp]
\centering
\includegraphics[width=12cm]{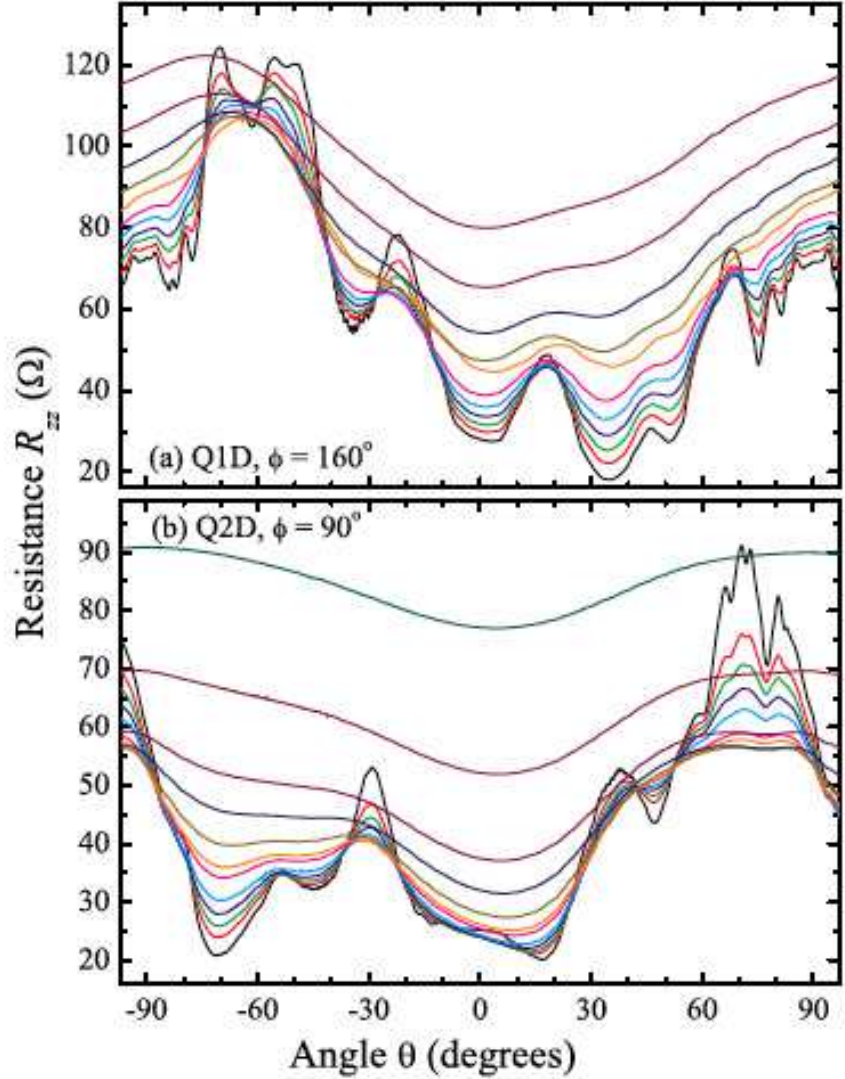}
\caption{Suppression of AMROs by increasing temperature
shown as interlayer resistance $R_{zz}(\propto \rho_{zz})$ of a
\cuscn ~crystal versus tilt angle $\theta$ for various
constant $T$ ($B=45$~T). (a)~Data for $\phi = 160^{\circ}$, a plane
of rotation at which $\rho_{zz}$ is determined by phenomena
on the quasi-one-dimensional Fermi-surface sections. In order of
increasing $R_{zz}$ at $\theta = 35^{\circ}$, the curves are for $T
= 5.3$, 6.5, 7.6, 8.6, 9.6, 10.6, 12.1, 13.1, 14.6, 17.1, 19.6 and
29.3~K respectively.
(b)~Similar data for $\phi=90^{\circ}$; here
$\rho_{zz}$ features are associated with the quasi-two-dimensional
Fermi-surface. In order of increasing $R_{zz}$ at $\theta = -70^{\circ}$,
the curves are for $T=5.3$, 7.6, 8.6, 9.6, 10.6, 12.1, 13.1, 14.6,
17.1, 19.6, 24.5 and 29.3~K respectively.}
\label{hightamro}
\end{figure}

Finally, note that increasing the temperature 
gradually suppresses the AMROs 
(see Figures~\ref{hightamro}(a) and (b)).
Modelling of this phenomenon shows that it can be described
by a temperature-dependent scattering rate,
$\tau^{-1}=\zeta + \chi T^2$, where $\zeta$
and $\chi$ are constants~\cite{goddardnew}.
The exponent suggests that the suppression
of AMROs is due to electron-electron scattering.
Another interesting feature of this suppresion
is that the same scattering rate appears to
apply to orbits on the quasi-one-dimensional
and quasi-two-dimensional parts of the Fermi surface.
This suggests 
that mechanisms for superconductivity
in organic metals
that invoke a large variation
in scattering rate over the FS ({\it e.g.}
``FLEX'' methods~\cite{schmalian})
may be inappropriate for \cuscn.

\subsection{Further clues about dimensionality in the 
resistivity tensor components}
\label{s1p3p3}
In the past there has been some confusion as to the
origin of AMROs in quasi-two-dimensional charge-transfer salts
of BEDT-TTF; and in particular, the component
of the resistivity tensor in which the oscillations occur.
There are few reliable measurements of the in-plane
conductivity or resistivity $\rho_{||}$
of quasi-two-dimensional
crystalline organic metals~\cite{review,r1};
experiments involving conventional edge contacts are
problematic~\cite{review,r1}. Because of the very
large resistivity anisotropy, such data are almost
often dominated by the much larger interplane
resistivity component, $\rho_{zz}$~\cite{review}.
In order to circumvent
this problem, a number of authors
have turned to a MHz skin depth
technique to measure the field dependence of the
in-plane resistivity~\cite{r1,mielke,frohlich,coffey};
this technique is very suitable for pulsed 
magnetic fields~\cite{r1,mielke,coffey}.
Samples for such experiments
are mounted within a small coil
which forms part of the tank circuit of a
tunnel-diode oscillator; shifts in frequency can
be related to changes in the skin depth and hence
to changes in the in-plane conductivity~\cite{mielke,frohlich}.
In some experiments~\cite{r1,frohlich},
top and bottom contacts are also mounted
on the sample, so that simultaneous measurements
of the interplane resistivity component, $\rho_{zz}$
can be made.

\begin{figure}[htbp]
\centering
\includegraphics[width=9cm]{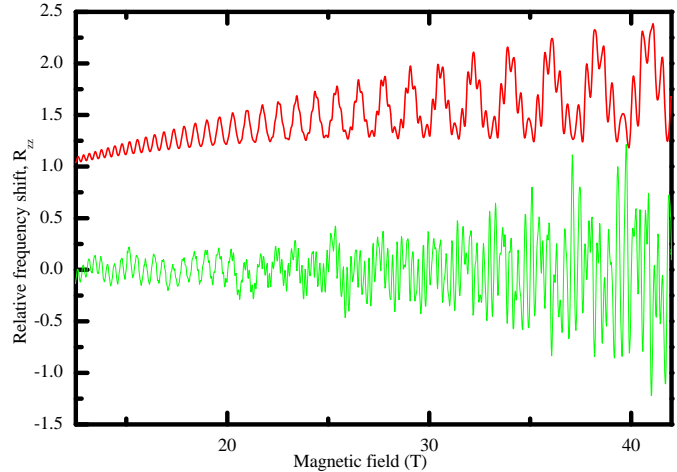}
\caption{Simultaneous measurement of $\rho_{zz}$ (upper trace)
and the frequency shift of a tunnel diode oscillator
(related to the in-plane component of the resistivity) (lower trace).
The sample is a single crystal of \cuscn; the temperature is 470~mK.
Note that the rapid magnetic quantum oscillations due to magnetic
breakdown are much more prominent in the lower data set.
(After Ref.~\cite{edwardssingleton}.)}
\label{freq1}
\end{figure}

Figure~\ref{freq1} shows a comparison of $\rho_{zz}$
and the frequency shift,
measured simultaneously. The most noticeable contrast between the
two data sets is the much more prominent magnetic
breakdown (the higher frequency) oscillations 
in the (in-plane) frequency data.
Although a quantitative model of such oscillations
poses some theoretical difficulties, it is easy to see
qualitatively why magnetic breakdown will influence
the in-plane conductivity much more than it does the
interplane conductivity. Magnetic breakdown represents
the tunnelling of quasiparticles from the one
Fermi-surface section to another~\cite{harrisonbd}.
This will affect the way in which a quasiparticle's
velocity evolves with time, and hence the conductivity.
As virtually all of the dispersion of the electronic bands
is in-plane, a magnetic breakdown event will have a relatively
large effect on the time dependence of the in-plane
component of a quasiparticle's velocity.
By contrast, the warpings in the interplane direction
of both sections of the the Fermi surface of
\cuscn ~seem to be rather similar~\cite{goddard};
hence magnetic breakdown will have comparatively
little effect on the time evolution of the
interplane component of the quasiparticle velocity.

\begin{figure}[hbp]
\includegraphics[width=11cm]{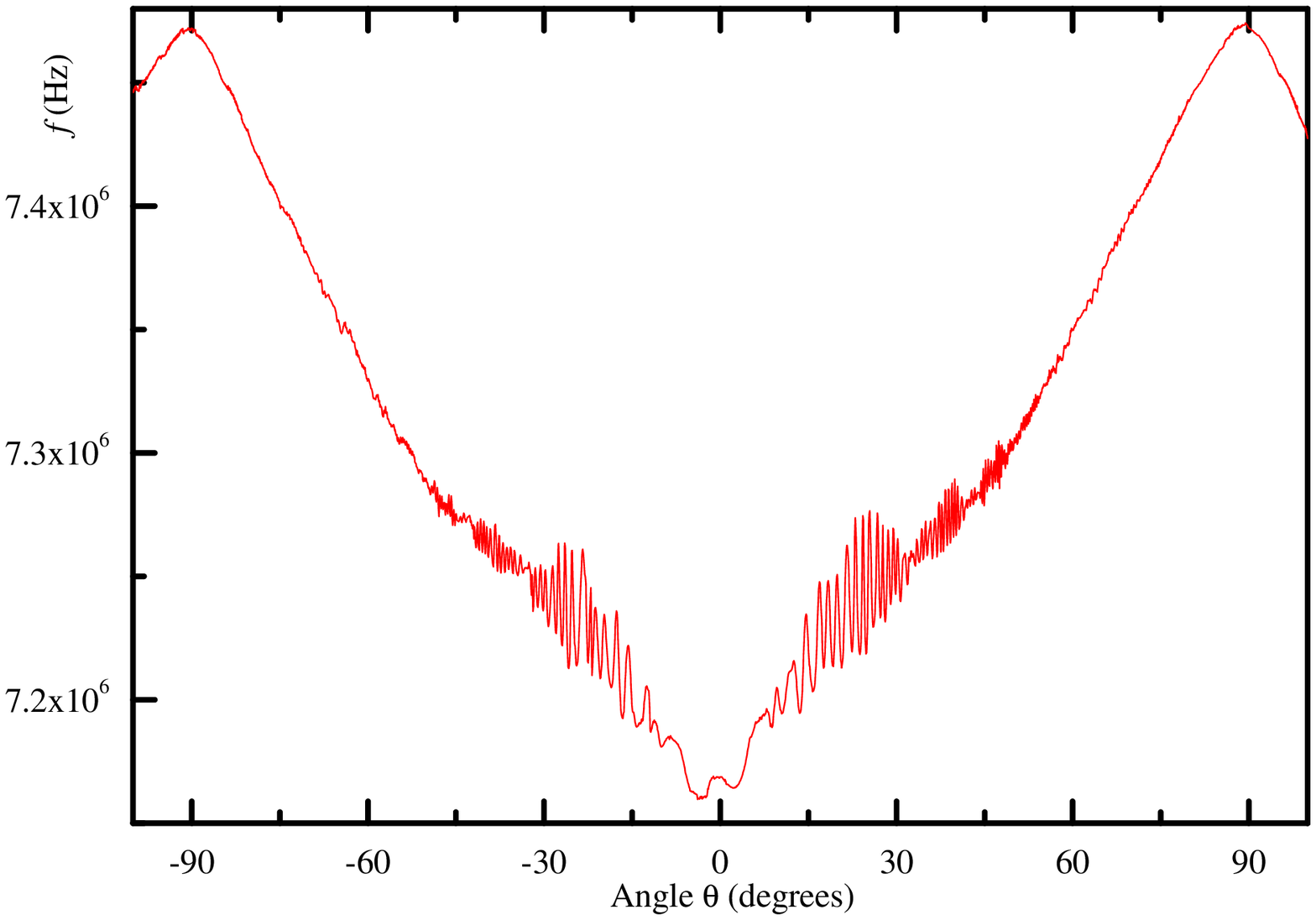}
\includegraphics[width=11cm]{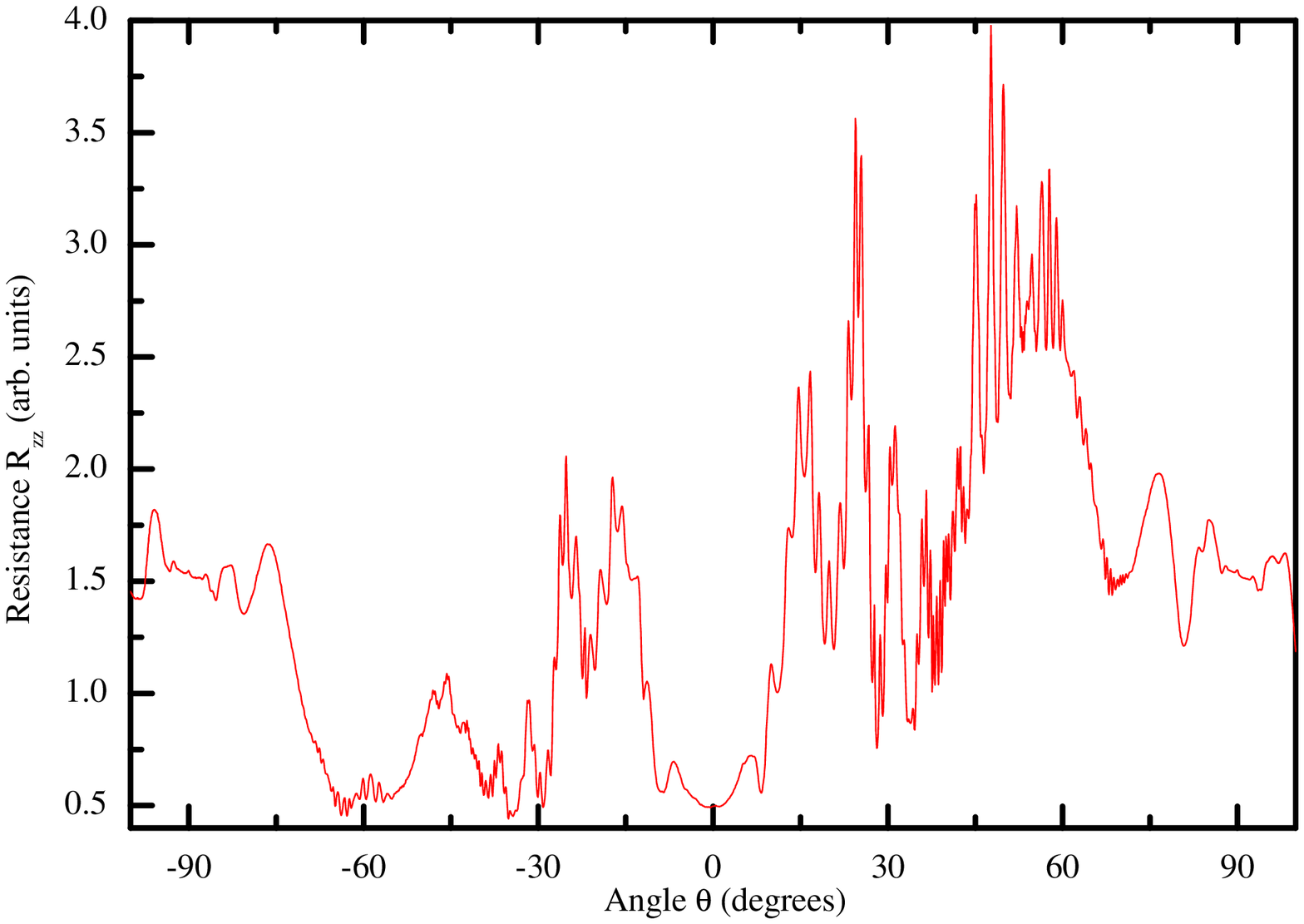}
\caption{Comparison of the magnetic-field orientation dependence of
the interlayer resistance $R_{zz}$ (lower trace-
proportional to $\rho_{zz}$)
and that of the frequency shift of a tunnel diode
oscillator (upper trace- related to $\rho_{||}$)
measured simultaneously. $\theta = 0$ corresponds to the field
being normal to the quasi-two-dimensional planes of
\cuscn . The temperature is 470~mK and the static magnetic field is 42~T.
The rapid oscillations close to $\theta =0$ in both figures are
Shubnikov-de Haas oscillations. The slower oscillations,
periodic in $\tan \theta$, and only seen in the lower trace, are AMROs.
The peak denoting interlayer coherence is visible at $\theta=90^{\circ}$
in the lower figure
(after Reference~\cite{edwardssingleton}).}
\label{freq2}
\end{figure}

Figure~\ref{freq2} contrasts the angle dependence of the
two components of the resistivity at fixed magnetic field.
Whereas $\rho_{zz}$ exhibits
strong AMROs, the frequency (depending the in-plane resistivity)
shows none. This is entirely consistent with the
expectations of semiclassical models of AMROs~\cite{amro,goddardprb2004}.

It should be noted that a high-frequency variant of 
AMROs, known as the {\it Fermi-surface-traversal
resonance} (FTR), has been developed. This 
high-frequency (GHz) magneto-optical 
technique allows additional information about
the topology and corrugations of quasi-one-dimensional 
Fermi sheets to be deduced~\cite{review,schrama,kovalev} .

\section{High-field Shubnikov-de Haas measurements and 
quasiparticle scattering. }
\label{s1p4}
\label{The effect of reduced dimensionality}
\label{s1p4p1}
Above we saw that the effect of the small but finite
interlayer transfer integral on the Fermi-surface topology
is very important in phenomena
such as angle-dependent magnetoresistance oscillations (AMROs).
However, from the perspective of magnetic quantum oscillatory phenomena such
as the Shubnikov-de Haas and de Haas-van Alphen effects,
the Fermi surface properties behave in an almost ideally two-dimensional
way. To see how this occurs, 
consider the Landau quantization of
the quasiparticle states due to a 
magnetic field {\bf B}:~\cite{singlebook,ashcroft}
\begin{equation}
E({\bf B},k_z,l)=\frac{\hbar e |{\bf B}|}{m^*}(l+\frac{1}{2})
+E(k_z) \equiv \hbar\omega_{\rm c}(l+\frac{1}{2}) +E(k_z).
\label{landau}
\end{equation}
Here $E(k_z)$ is the energy of the (unmodified) motion
parallel to {\bf B}, $l$ is the Landau quantum number ($0,1,2,\dots$)
and $m^*$ is an orbitally-averaged effective mass; 
the angular frequency $\omega_{\rm c}= eB/m^*$ 
(the {\it cyclotron frequency})
corresponds to the semiclassical
frequency at which the quasiparticles orbit
the Fermi surface~\cite{singlebook,ashcroft}.
(For the moment
we neglect the Zeeman term due to spin~\cite{review}.)
In virtually all practical experiments, the magnetic field is
perpendicular to the highly-conducting planes or tilted by angles
less than $60^{\circ}$ from this direction.
In such situations, $E(k_z)$ will be $\sim t_{\perp}$;
for most of the quasi-two-dimensional charge-transfer salts,
the Landau-level energy spacing will eclipse $t_{\perp}$
in fields of order $1-5$~T.
The necessity to use a rigorously two-dimensional
approach to analyse the Shubnikov-de Haas (resitivity)
and de Haas-van Alphen (magnetization)
oscillations~\cite{review,wosnitza,shoenberg} 
in these cases cannot
be overemphasized;
see References~\cite{oldharrison,harrisonbd} for a thorough 
discussion of this point.

\subsection{The deduction of quasiparticle scattering rates}
\label{s1p4p2}
Recently, it has been proposed that 
the dependence of superconducting
properties
on the quasiparticle scattering rate is 
an excellent way of identifying the mechanism
for superconductivity in quasi-two-simensional organics~\cite{r0a}.
Unfortunately, this is not as straightforward as it might seem.
A measure of the scattering rate in
metallic systems is often derived from
the rate at which magnetic quantum oscillations
(such as de Haas-van Alphen oscillations)
grow in amplitude with increasing magnetic field;
the dominant term in the Lifshitz-Kosevich formula
(see {\it e.g.}  Refs.~\cite{oldharrison,shoenberg})
describing this phenomenon is 
$\exp[-14.7m^*_{\rm CR}T_{\rm D}/B]$ (SI units).
The constant describing
the phase-smearing of the oscillations
due to Landau-level broadening is 
the so-called {\it Dingle temperature}, $T_{\rm D}$. 

If one were to assume that $T_{\rm D}$
is due {\it solely} to scattering 
({\it i.e.} the energy width of the
Landau levels detected by the de Haas-van Alphen effect
is associated only with
their finite lifetime due to scattering) then 
$T_{\rm D}$ would be related to the scattering rate
$\tau_{\rm dHvA}^{-1}$ by
the expression
\begin{equation}
T_{\rm D}= \frac{\hbar}{2 \pi k_{\rm B} \tau_{\rm dHvA}}.
\label{dinglebong}
\end{equation}
In many experimental works, it is assumed
that the $\tau_{\rm dHvA}$ deduced from 
$T_{\rm D}$ in this way is a true measure of
scattering rate; however, in quasi-two-dimensional
organic metals this is almost certainly not the case.
The problem becomes apparent if such 
scattering times are compared with the
$\tau_{\rm CR}$ deduced from cyclotron 
resonance experiments~\cite{edwardssingleton}.
In such cases the following inequality is found;
\begin{equation}
\tau_{\rm CR} > \tau_{\rm dHvA}.
\end{equation}
For some layered metals (see, for example, 
the work of Hill~\cite{hillscatt}), 
the $\tau_{\rm CR}$
measured in cyclotron resonance has been
found to be four to ten times larger than $\tau_{\rm dHvA}$;
An example of this is shown in Figure~11
of Ref.~\cite{edwardssingleton},
where the insertion of the scattering rate inferred from
Shubnikov-de Haas oscillations
into a model for the cyclotron resonance produces linewidths
that are much too broad. A more realistic linewidth
is obtained with a longer scattering time.
As we shall now describe,
spatial inhomogeneities are the likley culprit 
for the difference of scattering times.

\begin{figure}[tbp]
   \centering
\includegraphics[height=6cm]{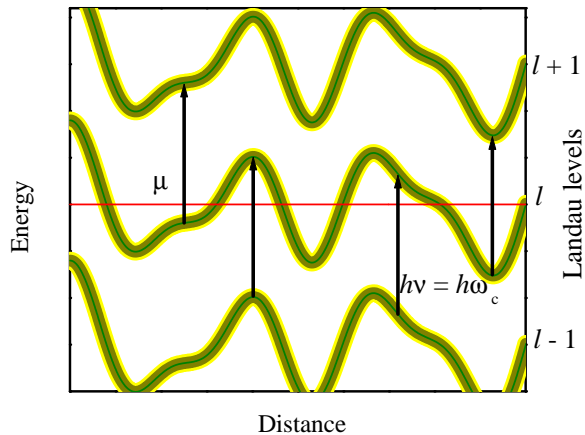}
\caption{Cartoon of the effect of spatial inhomogeneities
on Landau-level widths and energies. 
The variations in the potential experienced
by the electrons make the Landau levels (shaded curves)
move up and down
in energy as one moves through the sample. 
As the field is swept, the levels
will move up through the chemical potential $\mu$
and depopulate, resulting in the de Haas-van Alphen
and Shubnikov-de Haas effects.
The Dingle Temperature essentially
parameterizes the movement  of the
{\it total energy width of a
Landau level} through $\mu$;
hence it will measure a width that
includes the energy variation due to inhomogeneities.
By contrast, cyclotron resonance (shown by
vertical arrows) is a ``vertical'' transition;
it will measure just the true
width of the Landau levels due to scattering (represented by
shading).
}
\label{cartoon}
\end{figure}

Screening is less effective in
systems containing low densities of
quasiparticles (such as organic metals),
compared to that in elemental metals;
hence variations in the potential experienced by the
quasiparticles can lead to
a spatial variation of the Landau-level energies 
(see Figure~\ref{cartoon}).
Even in the (hypothetical) complete absence
of scattering, Harrison~\cite{nhandjs} has shown that this
spatial variation would give the Landau level a finite energy width
(see Figure~\ref{cartoon})
and therefore lead to an 
apparent Dingle temperature
\begin{equation}\label{dingle}
    T_{\rm D}=\frac{\bar{x}[1-\bar{x}]F^\prime(\bar{x})^2a}
    {\pi k_{\rm B}m^\ast}\sqrt{\frac{\hbar e^3}{2F}}.
\end{equation}
Here $F$ is the magnetic-quantum-oscillation frequency,
and $F^\prime = {\rm d}F/{\rm d}x$;
$x$ represents the (local) fractional
variation of the quasiparticle density due to the
potential fluctuations and $\bar{x}$ is
its mean.

The Dingle temperature measured in experiments
will therefore usually represent a combination of the
effects described by Equations~\ref{dinglebong}
and \ref{dingle}. Hence the simple-minded use
of Equation~\ref{dinglebong} to yield
$\tau_{\rm dHvA}$ from $T_{\rm D}$
will tend to result in
a parameter that is
an overestimate of the true 
scattering rate (see Figure~\ref{cartoon}).
By contrast, cyclotron resonance (shown by
vertical arrows in Figure~\ref{cartoon}) 
is a ``vertical'' transition (due to the
very low momentum of the photon);
it will measure just the true
width of the Landau levels due 
to scattering (represented by
shading)~\cite{edwardssingleton}.

Providing a thorough  treatment of the 
sample's high-frequency magnetoconductivity is
made~\cite{edwardssingleton}, then the measured scattering rate
deduced from a cyclotron 
resonance experiment is 
a good measure of the energy width of
the Landau levels due to their finite lifetime.
Once this has been realized, 
cyclotron resonance can be used to
give quantitative details of the
quasiparticle scattering mechanisms.
By contrast, the {\it apparent} scattering rate 
deduced from de Haas-van Alphen
and Shubnikov-de Haas oscillations 
can contain very significant contributions
from spatial inhomogeneities~\cite{nhandjs}.

Finally, one should add that the 
scattering rate $\tau_{\sigma}^{-1}$
measured in
a zero-field {\it in-plane}
conductivity experiment can be very
significantly different from $\tau_{\rm CR}^{-1}$
because the two measurements are sensitive to
different types of scattering process.
One of us has discussed this issue in 
detail in two recent papers~\cite{CR,condprob};
in particular, in the case of the
$\kappa$-phase BEDT-TTF salts,
there is an (as-yet) unexplained
quatitative discrepancy between the size of the
in-plane conductivity
and other measures of the quasiparticle properties~\cite{condprob}.
Further work to resolve this question is necessary.

Finally, ``jitter'' in the Brillouin zone boundary may be yet 
another source of scattering. This is predicted 
to give rise to 'hot spots' where the 
Fermi surface intersects the Brillouin zone boundary~\cite{izabela}. 

\section{Charge-density waves at fields above 
the Pauli paramagnetic limit}
\label{s1p5}
Intense magnetic fields ($B$)
impose severe constraints 
on spin-singlet paired electron states. 
Superconductivity is one example of a 
groundstate where this is true, although 
orbital diamagnetic effects usually destroy 
superconductivity at lower magnetic fields 
than does the Zeeman effect~\cite{r11,r12}. 
Charge-density wave (CDW) systems, by comparison, 
are mostly free from orbital effects~\cite{r13}, 
and so can only be destroyed by coupling 
$B$ directly to the electron spin. 
While most CDW systems have gaps that are 
too large to be destroyed in laboratory-accessible 
fields~\cite{r13}, several new compounds 
have been identified within the last decade 
that have gaps ($2\Psi_0$) as low as a few meV, 
bringing them within range of 
state-of-the-art static magnetic 
fields.

As we shall discuss in Section~\ref{s1p6},
\mhg (where $M = $K, Tl or Rb; $2\Psi_0 \sim 4$~meV) 
is one example that has been extensively studied~\cite{r14}. 
However, it has a complicated phase diagram in a 
magnetic field owing to the imperfect nature of 
the nesting~\cite{r15}; closed orbits exist 
after the Fermi-surface reconstruction 
which become subject to Landau quantization in a 
magnetic field~\cite{r16}, potentially modifying 
the groundstate. By contrast, 
\per ~(where M = Pt and Au) appears to be 
fully gapped~\cite{r17}.
However, the existence of spin $\frac{1}{2}$
moments on the Pt sites makes the $M =$Au 
system a pristine example of a small-gap, fully 
dielectric CDW material. 

Measurements of the CDW transition temperature $T_{\rm P}$
(where the subscript ``P'' stands for ``Peierls'')
in \pera ~
($T_{\rm P} = 11$~K at $B=0$) as a function of 
$B$ indicate that it is suppressed in a 
predictable fashion~\cite{r18}, 
allowing a Pauli paramagnetic limit
of $B_{\rm P} \approx 37$~T to be inferred.
However, the closure of the CDW gap with field is 
in fact considerably more subtle~\cite{r19};
a finite transfer integral $t_a$ perpendicular to the
nesting vector produces a situation analogous to
that in an indirect-gap semiconductor, where
the minimum energy of the empty states above the chemical
potential $\mu$ is displaced in $k$-space
from the maximum-energy occupied states
below $\mu$~\cite{r19}.
Consequently, Landau quantization
of the states above and below $\mu$
is possible, leading to a
thermodynamic energy gap $E_{\rm g}(B,T)$ of
the form~\cite{r19}
\begin{equation}
\label{gapfield}
E_{\rm g}(B,T)=2\Psi(T)-4t_a-g\mu_{\rm B}B
+\gamma\hbar\omega_{\rm c}.
\end{equation}
Here, $\Psi(T)$ is the temperature-dependent
CDW order parameter ($\Psi(T) \rightarrow \Psi_0$ 
as $T \rightarrow 0$), 
$\omega_{\rm c}$ is a characteristic cyclotron frequency
in the limit $B\rightarrow 0$, and $\gamma$
is a nonparabolicity factor;
$g\approx 2$ is the Land\'{e} g-factor
and $\mu_{\rm B}$ is the Bohr magneton.
Note that the Landau quantization
{\it competes} with 
Zeeman splitting; however, at sufficiently
high $B$ it becomes impossible to sustain closed
orbits, leading to a straightforward dominance of the
Zeeman term~\cite{r19}.

\begin{figure}[htbp]
\centering
\includegraphics[width=8cm]{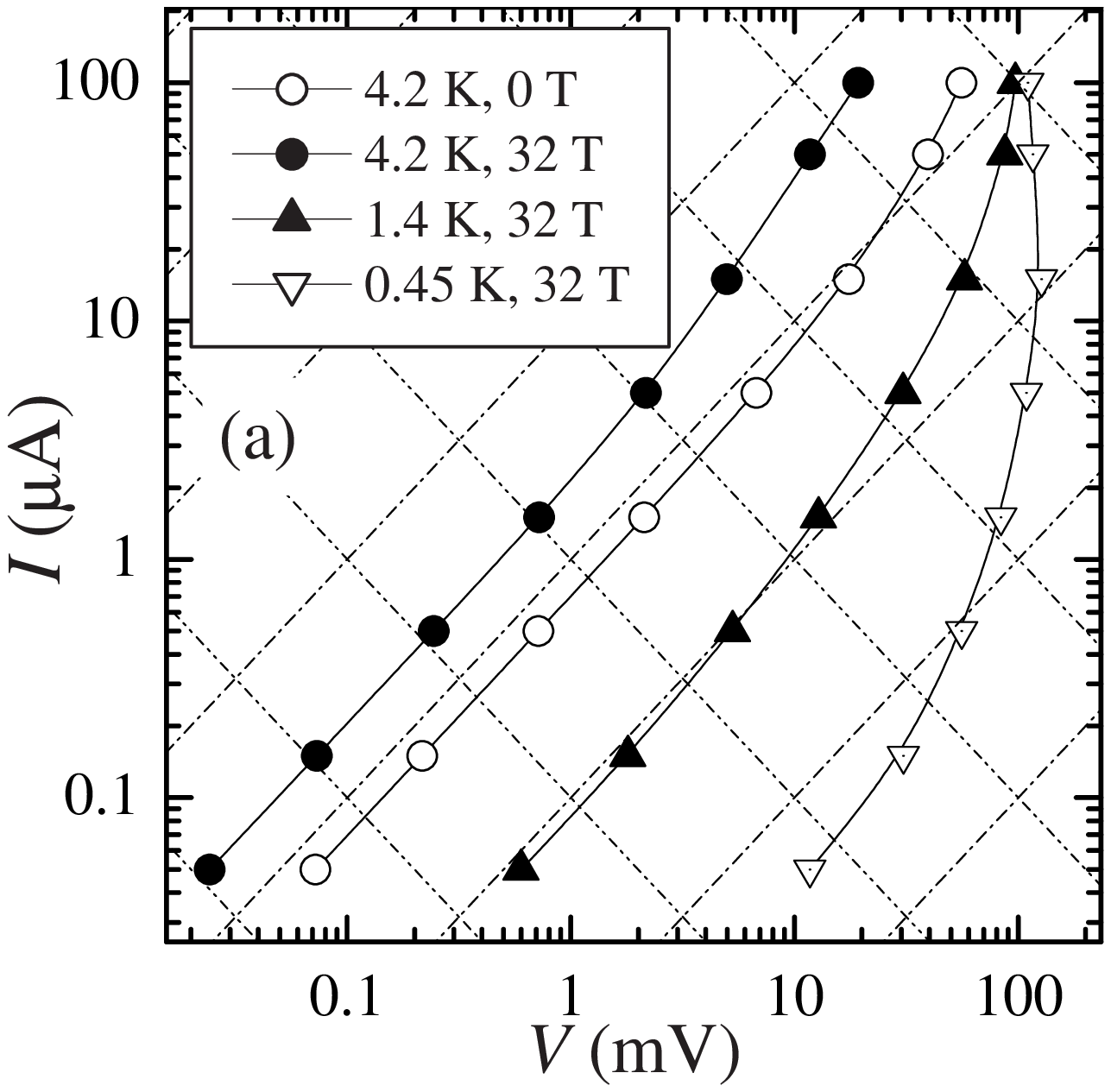}
\includegraphics[width=8cm]{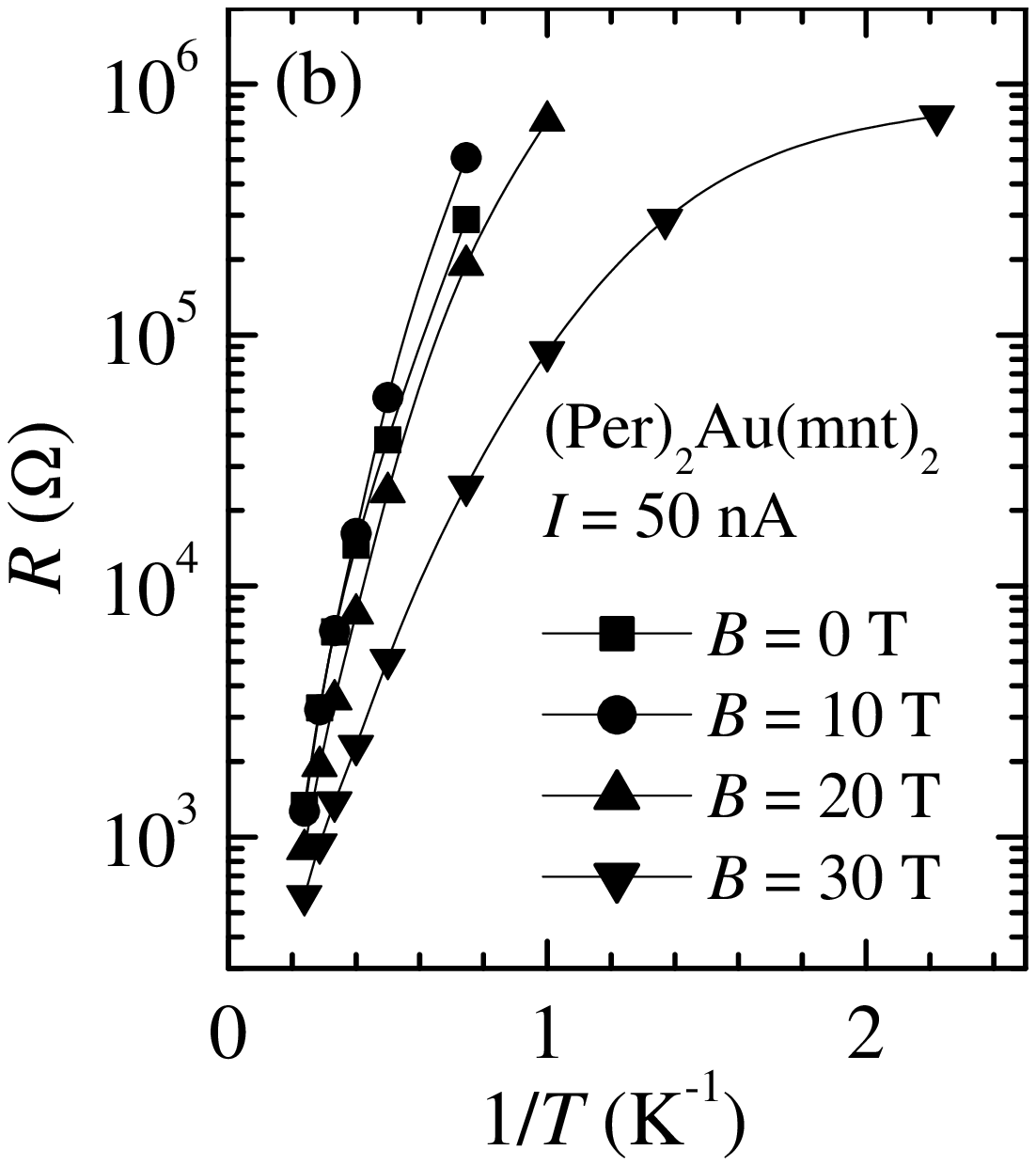}
\caption{(a)~Non-linear current-versus-voltage 
characteristics of \pera ~
plotted on a logarithmic scale for various
temperatures and fields (see inset key).
The negative-slope diagonal lines are contours 
of constant power and the 
positive-slope diagonal lines are contours of 
constant resistance, providing a guide 
as to when the sample's behavior is dominated by ohmic,
thermally-activated conduction rather than sliding.
(b)~Arrhenius plots of resistance $R$ (
$\propto \rho_{yy}$, logarithmic scale)
versus $1/T$ with $I=$~50~nA 
for \pera ~at several different
$B$ (after Ref.~\cite{r19}).}
\label{fig2}
\end{figure}

Another subtlety in assessing the field-dependent 
thermodynamic gap (and hence the Pauli
paramagnetic limit) in \pera ~stems from
the complicated nature of the low-temperature 
electrical conductivity, which contains contributions
from both the sliding collective mode of the CDW
and thermal excitation across the gap 
(see Fig.~\ref{fig2}(a))~\cite{r19,r20}.
This leads to a measured resistivity 
$\rho_{yy}\approx (\sigma_T+j_y/{E}_{\rm t})^{-1}$,
where $\sigma_T$ is the conductivity
due to thermal excitation across the
gap and $j_y/{E}_{\rm t}$ is the 
contribution from the collective mode,
$j_y$ being the current density and ${E}_{\rm t}$
the threshold electric field to depin the CDW. 
Now Eq.~\ref{gapfield}
contains $\Psi$, which is $T$-dependent;
moreover, ${E}_{\rm t}$ may also
depend on $T$.
Thus, Arrhenius plots are in general curved 
(see Fig.~\ref{fig2}(b)),
with a slope 
\begin{equation}\label{slope}
\frac{\partial\ln\rho_{yy}}{\partial (1/T)} \approx 
\frac{\frac{1}{2k_{\rm B}}(E_{\rm g}-T\frac{\partial E_{\rm  g}}{\partial T})
-\frac{j_yT^2}{\sigma_T{E}_{\rm t}^2}    
\frac{\partial {E}_{\rm t}}{\partial T}}
{1+\frac{j_y}{\sigma_T {E}_{\rm t}}}.
\end{equation}
With appropriate choice of temperature and bias regimes,
it is possible to make a reliable estimate
of $E_{\rm g}$ from plots such as those
in Fig.~\ref{fig2}(b). On the other
hand, it can be shown~\cite{r20a} that
poorly-chosen experimental conditions
can easily lead to errors in the
size of the derived gap.

Once accurate values of $E_{\rm g}(B)$
have been obtained, the method of
Ref.~\cite{r19} can be used to fit Eq.~\ref{gapfield}
by adjusting the parameters
$2\Psi_0+4t_a$, $4t_a$
and $v_{\rm F}$, where $v_{\rm F}$ is the Fermi velocity
in the metallic state;
in the CDW state it is used to parameterise
the quasiparticle dispersion.  
A good fit is obtained
using
$t_a=$~0.20~$\pm$~0.01~meV,
$v_{\rm F}=$~(1.70~$\pm$~0.05)~$\times 10^5$~ms$^{-1}$
and $2\Psi_0 =$~4.02~$\pm$~0.04~meV~\cite{r19}.
These parameters correspond to 
$E_{\rm g}=$~3.21~$\pm$~0.07~meV 
at $B=0$, $T=0$; the derived
transfer integrals
$t_a$ and $t_b~(\approx$~188~meV) are
in good agreement with theory~\cite{r20b}
and thermopower data~\cite{r17}.
(Note that these band parameters
exclude the possibility
of field-induced CDW (FICDW) states of the
kind proposed in Ref.~\cite{r20c} in \per ~salts.)

\begin{figure}[htbp]
\centering
\includegraphics[width=11cm]{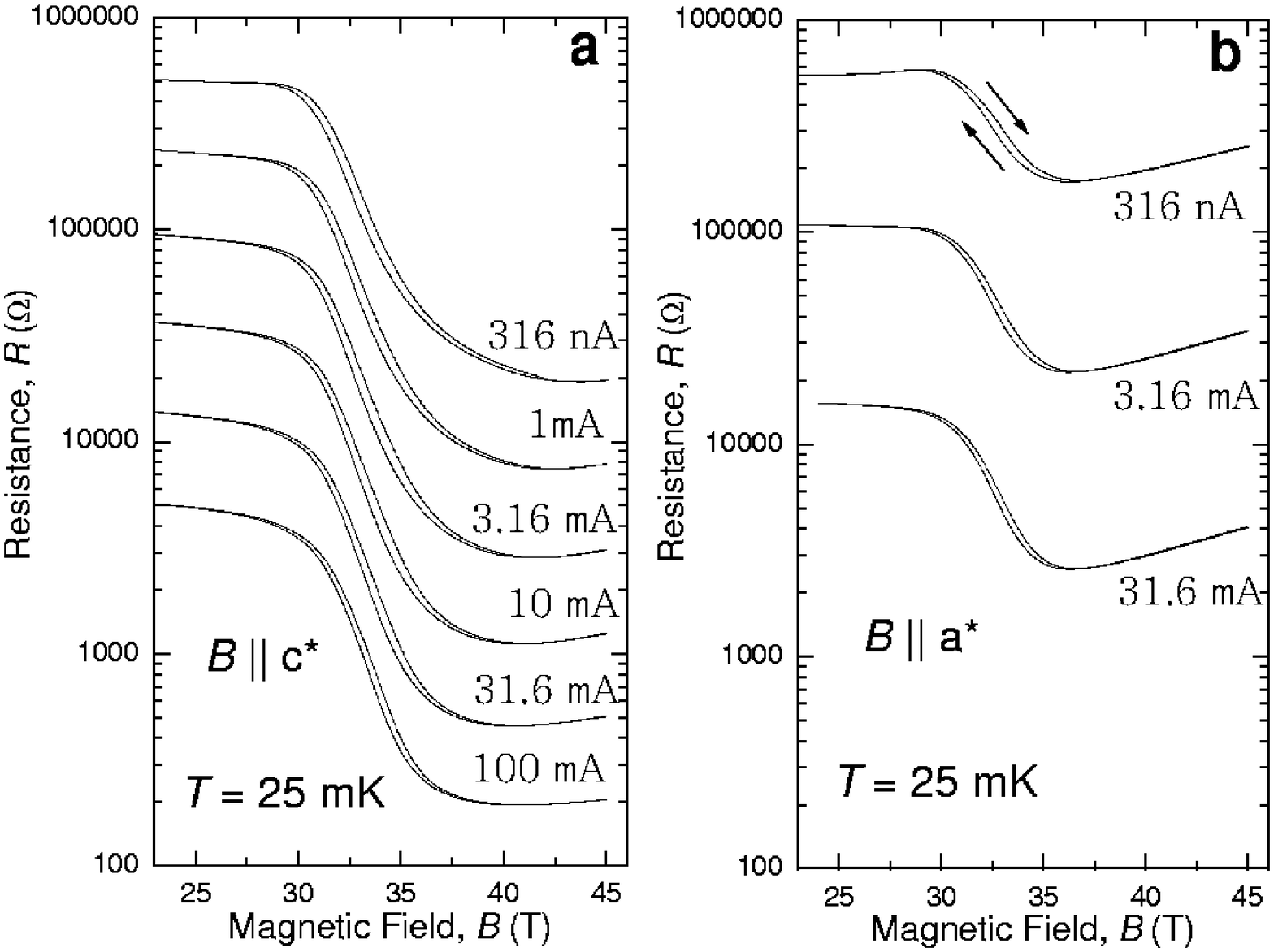}
\includegraphics[width=7.6cm]{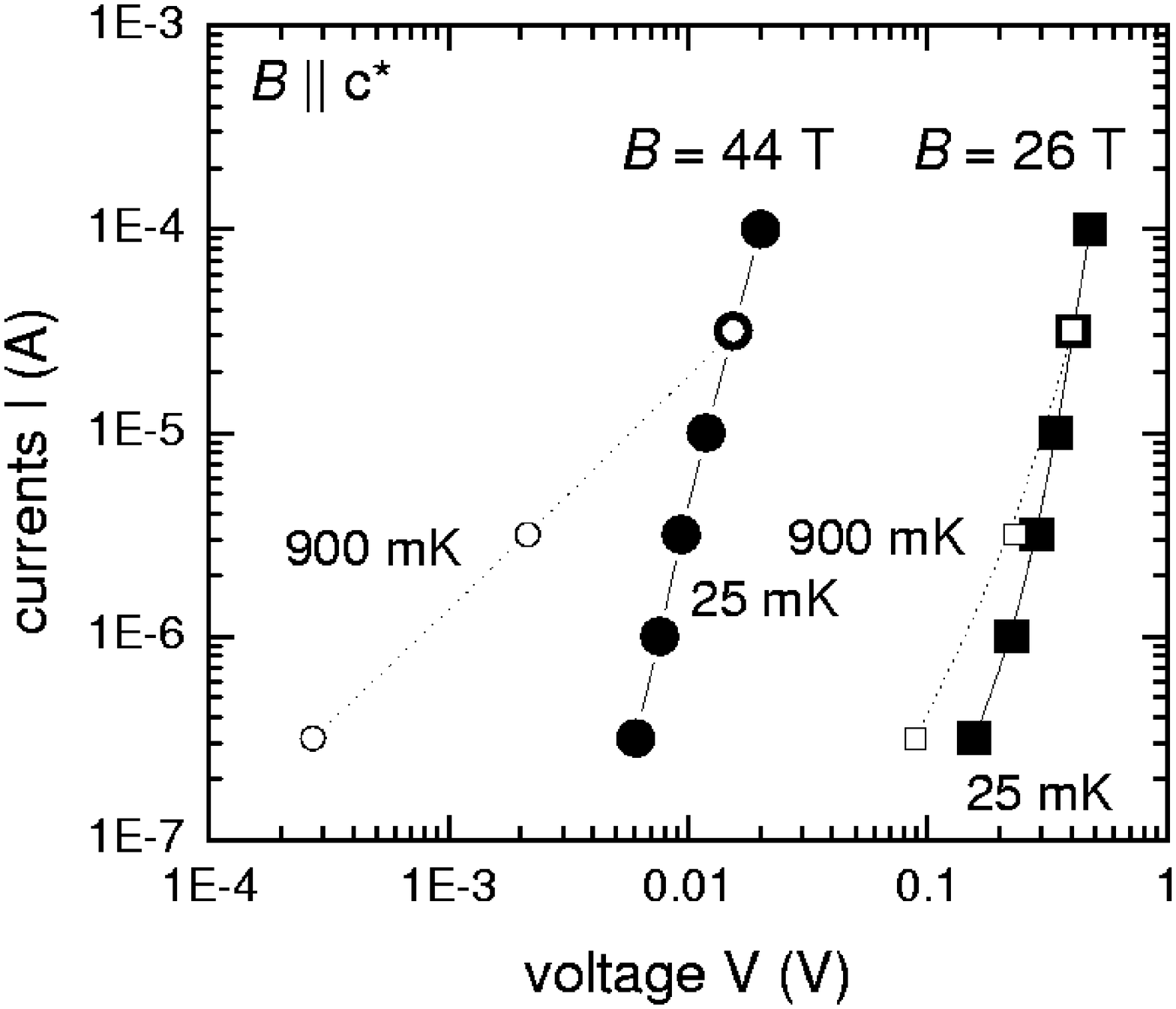}
\caption{Top: resistance of a single crystal of 
(Per)$_2$Au(mnt)$_2$
measured at 25~mK for
fields between 23 and 45~T, 
for two different orientations $c^\ast$ (a) and
$a^\ast$ (b) of $B$ perpendicular to its 
long axis $b$, at several different
applied currents.  
The lowest resistance for a given current occurs
for $B$ parallel to $c^\ast$, 
which is perpendicular to $a^\ast$. 
The dependence of
the resistance on current signals non-ohmic behaviour.
Bottom:~non-linear current-versus-voltage characteristic of
(Per)$_2$Au(mnt)$_2$ plotted on a log-log scale, 
for magnetic
fields (26 and 44~T) above (circles) and below (squares) $B_{\rm P}$. 
Filled symbols connected by solid 
lines represent data taken at 25~mK
while open symbols connected by 
dotted lines represent data taken at
900~mK. (After Ref.~\cite{r20}.)}
\label{fig3}
\end{figure}

Armed with the band parameters,
one obtains a reliable estimate
of the Pauli paramagnetic limit; 
$B_{\rm P}=(\Delta_0+2t_a)/\sqrt{2}gs\mu_{\rm B}\approx 30$~T.
This corresponds to a sharp drop in
measured resistance $R$ ($\propto \rho_{yy}$),
as shown in Fig.~\ref{fig3} (left side),
which displays data recorded at
$T=25$~mK.
At such temperatures, there are very
few thermally-activated 
quasiparticles indeed,
leaving only the CDW collective mode to conduct;
this gives rise to a $R$ 
in Fig.~\ref{fig3} 
that is strongly dependent on current.
On passing through $B_{\rm P}\approx 30$~T,
$R$ drops very sharply,
and there is also hysteresis between up- and
down-sweeps of the field.
The latter effect could be the consequence
of a first-order phase transition on reaching
$B_{\rm P}$, compounded by CDW pinning effects.  

However, the most interesting observation about
Fig.~\ref{fig3} is the fact that 
the strongly non-linear $I-V$ characteristics
persist at fields well above $B_{\rm P}$,
as can be seen from both
$R$ data and $I-V$
plots (right-hand side of
Fig.~\ref{fig3}).
One can conclude from these data that it is 
the CDW depinning voltage that changes at $B_{\rm P}$. 
This is probably a consequence of the 
CDW becoming incommensurate or of the order parameter 
of the charge modulation becoming considerably 
weakened~\cite{r13,r20}. Evidence for the latter is obtained 
by repeating the $I-V$ measurements at slightly higher 
temperatures of 900~mK (Fig.~\ref{fig3}, right side). 
This temperature is sufficient 
to restore Ohmic behaviour for $B > B_{\rm P}$, 
suggesting that a 
reduced gap for $B > B_{\rm P}$ 
allows quasiparticles to be more 
easily excited. 

There have been some interesting
proposals for FICDW phases (e.g. Ref.~\cite{r20c}).
These mechanisms require the system not to be 
completely gapped; instead possessing 
a closed orbit for one of the spins which would then have a Landau gap 
at the chemical potential. Such a situation typically leads to 
the quantum Hall effect, and a metallic  behaviour 
of longitudinal resistivivity~\cite{chaikin1}. 
Current would then be able to flow without the 
CDW having to be depinned.
These effects nevertheless appear to be absent
from the data of Fig.~\ref{fig3};
the continuation of the non-linear CDW electrodynamics 
for $B > B_{\rm P}$  suggests that 
metallic behaviour is not regained. Instead, both spin 
components of the Fermi surface are most likely to be gapped independently
with differing nesting vectors,
leading to an exotic CDW phase 
that has some analogies with the
FFLO state of superconductors. 
The presence of two distinct, spin-polarized
CDWs with different periodicities 
will furnish separate spin and charge modulations 
that could in principle be detected using 
a diffraction experiment~\cite{r20}.

It has been suggested~\cite{lebed1} that
FICDWs occur in charge-density wave (CDW) systems in
strong magnetic fields when orbital quantization 
facilitates nesting of quasi-one-dimensional Zeeman-split bands.  
The free energy is minimised at 
low integral Landau subband filling factors $\nu$ by
the formation of a Landau gap at the Fermi energy~\cite{lebed1}. 
Hence, as is the case in field-induced spin-density 
wave states (FISDW)~\cite{chaikin1}, orbital quantization 
is implicit in FICDW formation, yielding a Hall conductivity
$\sigma_{xy}\approx 2\nu e^2/ah$ (where $a\sim$~20~\AA~ is the layer
spacing) and a longitudinal conductivity
$\sigma_{xx}\sim\sigma_0\exp[-\Delta/k_{\bf B}T]$ that is very
small and thermally activated ($\sigma_{xx}\ll\sigma_{xy}$).
Inversion of the conductivity tensor yields
\begin{eqnarray}\label{FICDWorb}
    \rho_{xx}\approx 
(ah/2\nu e^2)^2\sigma_0{\rm e}^{-\frac{\Delta}{k_{\rm B}T}}
    \ll\rho_{xy}\approx ah/2\nu e~\{\nu\geq 1\}\\
    \rho_{xx}\approx(1/\sigma_0){\rm e}^{+\frac{\Delta}{k_{\rm B}T}}\gg ah/2e^2
    \gg\rho_{xy}\approx 0~~~~\{\nu=0\}.
\end{eqnarray}
Samples of (Per)$_2$Pt(mnt)$_2$ have the 
ideal topology (1000~$\mu$m~$\times$~50~$\mu$m~$\times$~25~$\mu$m) 
for deriving reliable estimates of 
$\rho_{xx}$ from the $R_{xx}$ data of Ref.~\cite{r20c}, 
yielding 130~$\lesssim\rho_{xx}\lesssim$~400000~$\mu\Omega$cm, 
greatly exceeding the maximum metallic resistivity 
$ah/2e^2\approx$~26~$\mu\Omega$cm throughout 
all of the proposed FICDW phases by as much as a 
factor of 20000. The data of Ref.~\cite{r20c} 
are therefore consistent with $\nu=$~0 throughout. 
For the quantized nesting model to apply, 
each FICDW state would have different values of $\nu$, 
only one of which can be 0.

\begin{figure}[htbp]
\centering
\includegraphics[width=10cm]{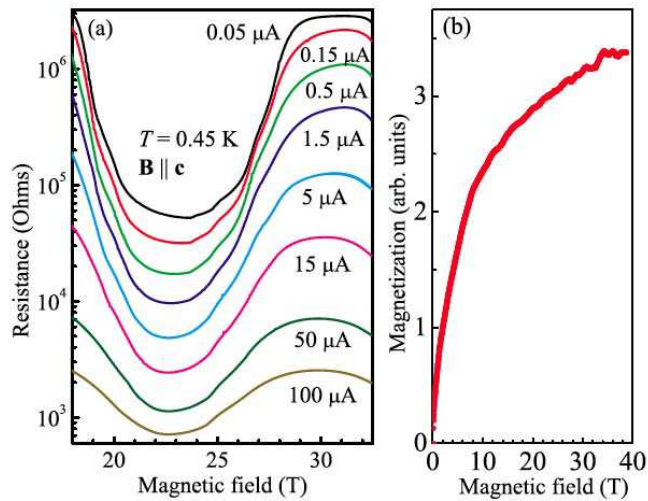}
\caption{(a)~Resistance versus field for 
(Per)$_2$Pt(mnt)$_2$ at various currents
measured by the present authors.
(b)~Magnetization $M$ of many
randomly oriented
(Per)$_2$Pt(mnt)$_2$ crystals at $T=0.50$~K;
$M$ does not saturate by $B \approx 40$~T~\cite{mcdonald1}.}
\label{fig11}
\end{figure}

Varying the current, $I$, causes
$R_{xx}$ and thus $\rho_{xx}$ 
vary considerably, reaching values in 
Fig.~\ref{fig1}a
that exceed $ah/2e^2$ by a factor $\approx$~10$^7$. 
Its strong dependence on the current density 
$j$ is consistent with a sliding 
collective mode contribution to $\sigma_{xx}$ 
(for all fields $B<$~33~T and $\nu=$~0), yielding 
\begin{equation}\label{CDW}
\rho_{xx}\approx
[\sigma_0 \exp-(\Delta/k_{\rm B}T)+j/E_{\rm t}]^{-1},
\end{equation}
where $E_{\rm t}$ is a threshold CDW depinning 
electric field~\cite{mcdonald1}, 
that may itself depend on $T$. 
Note that the near-linear $I$ versus 
voltage $V$-plots for $IV<$~2~$\mu$W (for different 
values of $B$, $T$ and $I$ in 
Fig.~2 of Ref.~\cite{mcdonald1}) suggests 
that heating is not a significant factor for 
$I<$~5~$\mu$A in Fig.~\ref{fig11}a. 

Hence, the behavior of the fully gapped \per 
system does not fit the usual definition 
of  ``magnetoresistance'' but is the
consequence of magnetic field-induced 
changes in the electric field $E_{\rm t}$ required to 
depin the CDW from the lattice, 
where $\nu=$~0  throughout.
Such behavior is inconsistent with the
quantized nesting model which requires different 
values of $\nu$ for each subphase~\cite{lebed1}; thus
the steps in
$\rho_{xx}$ probably correspond to field-induced changes in
$E_{\rm t}$.

The cooperative dimerization of 
the Pt spins in (Per)$_2$Pt(mnt)$_2$
can easily provide a mechanism for additional phase transitions or
changes in $E_{\rm t}$ compared to
(Per)$_2$Au(mnt)$_{2}$.  
The Pt spins couple strongly
to both the CDW, via distortions of the crystal lattice,
and $B$, as shown by the 
fact that they dominate the total
magnetic susceptibility 
(Fig.~\ref{fig11}(b)).  
Their effect on the
phase diagram is likely to be significant until all spins are fully
aligned by a field $B\gtrsim$~40~T (Fig.~\ref{fig1}(b)).

\section{A new quantum fluid in strong magnetic fields with 
orbital flux quantization}
\label{s1p6}
\khg ~is undoubtedly one of the most 
intriguing of BEDT-TTF-based charge-transfer 
salts~\cite{r14,r15,r16}. 
Like many other such materials, it possesses 
both two-dimensional (2D) and one-dimensional (1D) 
Fermi-surface (FS) sections. 
However, the 1D sheets are unstable at 
low temperatures, causing a structural phase 
transformation below $T_{\rm P} = 8$~K into a CDW state~\cite{chaikin1}.
Imperfect nesting combined with the continued existence 
of the 2D hole FS pocket gives rise to 
complicated magnetoresistance and 
unusual quantum-oscillation 
spectra at low magnetic fields and low temperatures~\cite{review}. 
At high fields, the CDW undergoes a number of 
transformations into new phases, many of which 
have been suggested to be 
field-induced CDW phases~\cite{r15}.

Undoubtedly the most exotic aspect of this material is 
its transformation into an unusual CDW state
above a characteristic field $B_{\rm k} = 23$~T 
(known as the ``kink'' transition); 
$B_{\rm k}$ is now known to correspond to the 
CDW Pauli paramagnetic limit~\cite{review,r16}, like \per. 
Such a regime is reached in \khg ~owing to the 
unusually low value  of $T_{\rm P}$~\cite{r16}. 
At fields higher than $B_{\rm k}$, Zeeman splitting 
of the energy bands makes a conventional CDW 
ground state energetically unfavourable~\cite{r24}, 
possibly yielding a novel modulated CDW state like 
that proposed for \pera ~(see Section~2
and Refs.~\cite{r19,r20}). 
In \khg ~this state is especially unusual due to 
the existence  of the 2D pocket, which appears 
to be ungapped by CDW formation~\cite{r16}. The CDW 
and 2D hole pocket screen each other, 
with pinning of the CDW then enabling a 
non-equilibrium distribution of orbital 
magnetization throughout the bulk~\cite{r16}. 
Consequently,
such a state exhibits a critical state analogous 
to that in type II superconductors.
Some of us have discussed this in more detail
in another Chapter of this book.

\section{Summary}
\label{s1p7}
\begin{figure}[htbp]
\centering
\includegraphics[width=9cm]{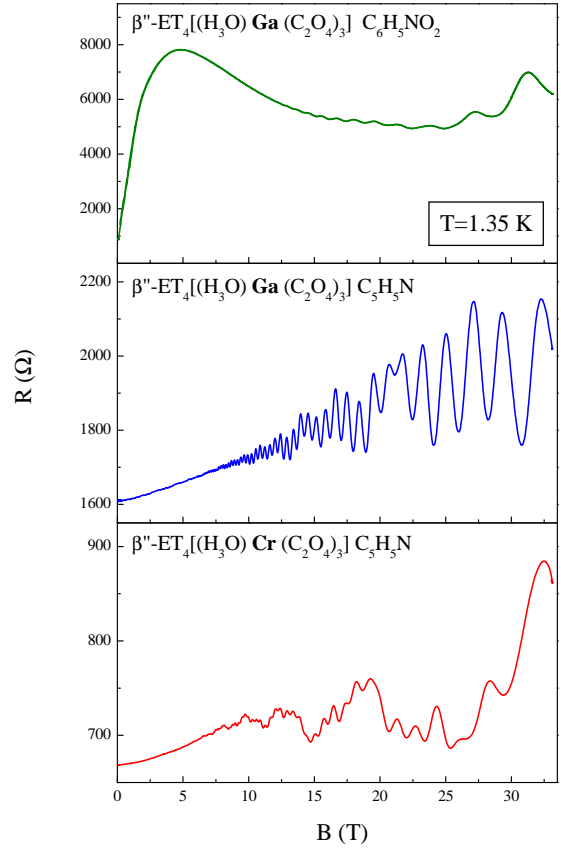}
\caption{Interlayer resistance of three different
charge-transfer salts of the form
$\beta"$-(BEDT-TTF)$_2$[H$_3$OM(C$_2$O$_4$)$_3$] (Sol)
(ET is an abreviation of BEDT-TTF).
The magnetic field is perpendicular to the highly-conducting
planes.
(after Reference~\cite{r7,r9a}).
}
\label{amalia}
\end{figure}

We have tried to illustrate why some members of the
high-field community are fond of charge-transfer salts.
This enthusiasm is likely to continue for some time, as
the charge-transfer salts offer great versatility
as a plaything for studying the formation
of bandstructure.
Using the known self-organisational properties
of small organic molecules, one can really
indulge in ``molecular architecture'',
in which the structure of a charge-transfer
salt is adjusted to optimise a desired property~\cite{mori}.
The most imaginative essays in this field involve
the use of molecules that introduce a further
property which modifies the electronic
behaviour, such as chirality or the presence
of magnetic ions~\cite{recentday}; in the latter
case, one of the aims is to manufacture an
organic Kondo system.
Some recent experimental data from three such compounds
are presented in Figure~\ref{amalia};
the salts in question have the generic formula
$\beta"$-(BEDT-TTF)$_2$[H$_3$OM(C$_2$O$_4$)$_3$] (Sol),
where M is a transition ion, and Sol is an incorporated solvent
molecule. The M ion allows one to introduce magnetic moments
in a controllable way, whereas changing the solvent molecule allows
fine details of the unit cell structure to be altered~\cite{recentday}.
As Figure~\ref{amalia} shows, such adjustments cause
very distinct changes to the Fermi-surface topology,
reflected in the magnetic quantum
oscillation spectra~\cite{r7,r9a}. Some of 
these salts appear to be superconductors. 

There are also many reasons for continuing to study charge-transfer salts
at high magnetic fields.
A particular goal is the ultraquantum limit,
in which only one quantised Landau-level is occupied;
phenomena such as yet more varieties
of field-induced superconductivity
have been predicted to occur once
such a condition is attained~\cite{r1,review,r9a}.
Furthermore, there are many open questions as to the
role of chiral Fermi liquids in such fields~\cite{review}.
Another area of considerable interest is the observation
of magnetic breakdown.
At fields above
50~T, the magnetic energy of the holes
in the organic superconductors
is starting to become a substantial
fraction of their total energy,
and one gradually starts to approach
the famous Hofstadter ``butterfly''
limit~\cite{review}.

\section*{Acknowledgements}
This work is funded by US Department of Energy 
(DoE) grant LDRD 20040326ER. Work at NHMFL 
is performed under the auspices of the 
National Science Foundation, DoE and the 
State of Florida.

\newpage


%
%
%

%

\begin{thebibliography}{99.}
\bibitem{footnote}
The existence of such a modulation might suggest that a beating
between ``neck and belly'' frequencies would be observed in the
dHvA effect~\cite{review}; however, in careful low-field studies~\cite{sasaki},
no such beating has been observed.
We shall see that this is because the typical cyclotron energy
is rather greater than $t_{\perp}$, even at the lowest fields used.
\bibitem{cokebottle}
When viewed in the extended zone scheme, the quasi-two-dimensional
section of this type of Fermi surface (blue in Fig.~\ref{belly}(c))
is often known as a ``coke-bottle'' Fermi surface (USA) or
a ``Cumberland sausage'' Fermi surface (Europe). (The latter
is a hand-made sausage with weak but approximately regular
corrugations.)
\end{thebibliography}
%



\printindex
\end{document}